\documentclass[aps,prd,twocolumn,superscriptaddress,nofootinbib,floatfix]{revtex4-2}

\usepackage[T1]{fontenc}
\usepackage{lmodern}
\usepackage{amsmath,amssymb,bm,graphicx}
\usepackage{microtype}
\usepackage{xcolor}
\usepackage{enumitem}
\usepackage[bookmarks=true]{hyperref}
\usepackage{orcidlink}

\hypersetup{
  colorlinks=true,
  linkcolor=blue,
  citecolor=blue,
  filecolor=blue,
  urlcolor=blue,
  pdftitle={Holography at finite resolution: zero-point length in AdS/CFT},
  pdfauthor={Ankit Anand, Kimet Jusufi, Amir A.~Khodahami, Adrinil Paul, Ahmad Sheykhi}
}

\newcommand{\dd}{\mathrm d}
\newcommand{\OO}{\mathcal O}
\newcommand{\PP}{\mathcal P}
\newcommand{\CC}{\mathcal C}
\newcommand{\lz}{l_0}
\newcommand{\intp}{\int\!\frac{\dd^3p}{(2\pi)^3}}

\begin{document}

\title{Holography at finite resolution: zero-point length in AdS/CFT}


\author{Ankit Anand \orcidlink{0000-0002-8832-3212}}
\email{ankitanandp94@gmail.com}
\affiliation{Department of
Physics, School of Basic Sciences, Indian Institute of Technology
Bhubaneswar, Bhubaneswar, Odisha 752050, India}

\author{Kimet Jusufi \orcidlink{0000-0003-0527-4177}}
\email{kimet.jusufi@unite.edu.mk} \affiliation{Physics Department,
University of Tetova, 1200 Tetova, North Macedonia}

\author{Amir A.~Khodahami \orcidlink{0000-0002-3327-3465}}
\email{a.khodahami@shirazu.ac.ir} \affiliation{Department of
Physics, College of Sciences, Shiraz University, Shiraz 71454,
Iran} \affiliation{Biruni Observatory, College of Sciences, Shiraz
University, Shiraz 71454, Iran}

\author{Adrinil Paul \orcidlink{0009-0005-0767-9778}}
\email{22ph05002@iitbbs.ac.in} \affiliation{Department of Physics,
School of Basic Sciences, Indian Institute of Technology
Bhubaneswar, Bhubaneswar, Odisha 752050, India}

\author{Ahmad Sheykhi \orcidlink{0000-0002-7496-4270}}
\email{asheykhi@shirazu.ac.ir} \affiliation{Department of Physics,
College of Sciences, Shiraz University, Shiraz 71454, Iran}
\affiliation{Biruni Observatory, College of Sciences, Shiraz
University, Shiraz 71454, Iran}


\begin{abstract}
We explore the zero-point length $l_0$ of T-duality-inspired
effective field theory in holography, distinguishing two readings.
As a boundary resolution scale, $l_0$ filters the source of a bulk
scalar in anti-de Sitter space and yields an exactly solvable
boundary response that is ultraviolet finite without a hard cutoff
and inherits the power-law tails of the zero-point-length kernel.
We show that this filter is the source-normalized bulk-to-boundary
transfer function of a $\nu=1$ scalar evaluated at radial depth
$z=l_0$. Anchoring Ryu-Takayanagi surfaces on the same radial
surface gives finite entanglement entropies that retain the
perimeter law and the universal sphere free energy. For two
intervals in AdS$_3$, the prescription yields the
mutual-information connectivity transition in closed form for
arbitrary interval lengths; equal intervals maximize the
connectivity window at fixed total length, and in the
subresolution regime the critical separation exhibits a universal
scaling law with a rescaled mutual information collapsing onto a
piecewise-linear profile. As an AdS-invariant bulk proper length,
$l_0$ renders the one-loop tadpole finite while preserving the
conformal form of boundary two-point functions. Its exact
K\"all\'en-Lehmann decomposition, however, has negative continuum
weight beginning at threshold, obstructing a unitary
interpretation of this propagator. The ratio $L/l_0$ identifies a
crossover scale of the form factor rather than a derived spectrum
of a unitary completion.
\end{abstract}


\maketitle

\section{Introduction}

The AdS/CFT correspondence relates gravitational observables in anti-de Sitter space to correlation functions of a conformal field theory on its boundary \cite{Maldacena,GKP,Witten}. In the scalar sector, the leading asymptotic coefficient of a bulk field acts as the source of a boundary operator, and the renormalized on-shell action generates its connected correlators \cite{deHaro,Skenderis}. Ultraviolet physics on the boundary is encoded in the near-boundary region of the bulk. In the standard treatment it is handled by a radial cutoff at $z=\epsilon$, which is removed by local counterterms. When the cutoff is instead kept at finite distance, the dictionary itself changes, as in the $T\bar T$ interpretation of cutoff AdS \cite{MMV,HKST,Kraus}. In both approaches the short-distance scale is introduced by hand rather than derived from an underlying minimal length.

A natural candidate for such a length is the \emph{zero-point length} $\lz$ \cite{Padmanabhan}. In the path-integral-duality formulation, and in its realization through string T-duality on compact directions \cite{SSP,FSP}, the massless propagator acquires a modified Bessel form factor that is exponentially soft in the ultraviolet. Moving this form factor from the propagator into a smeared source of a linear field equation produces regular static potentials and regular effective black-hole geometries \cite{Nicolini,Gaete,NicoliniReview}. The zero-point-length framework has also been extended to charged black holes \cite{Gaete:2022ukm}, with subsequent work investigating the Schwinger stability of extremal configurations inspired by T-duality \cite{Chen:2026nna}. The same mechanism has since been used to avoid the singularity and to establish geodesic completeness of T-duality black holes \cite{Jusufi:2026dky,JusufiNicolini,Jusufi:2026glp}, and to build modified cosmologies constrained by inflationary and late-time data \cite{LucianoSheykhi,LPS}. These applications live in asymptotically flat or cosmological spacetimes. To our knowledge, what a zero-point length means within AdS/CFT, where a short-distance scale has a precise boundary interpretation, has not been examined.

We find that the question has two distinct answers, depending on how $\lz$ enters the holographic dictionary. In the first reading, $\lz$ is a \emph{boundary resolution}: the source of a boundary operator is measured with the zero-point-length form factor. Our central object is
\begin{equation}
 W_{\lz}[J]\equiv W_0[\PP_{\lz}J],
 \qquad \PP_{\lz}=\lz\sqrt{-\partial^2}\,
 K_1\!\left(\lz\sqrt{-\partial^2}\right),
 \label{eq:definition}
\end{equation}
where $W_0$ is a fixed renormalized generating functional and $\lz>0$ is a boundary coordinate length. The operator is defined spectrally on the flat Euclidean boundary. Equation~\eqref{eq:definition} defines a finite-resolution source experiment, and it becomes a holographic prescription whenever $W_0$ has a bulk description. Source smearing as such is not new; what is specific here is the Bessel filter, whose non-analytic small-momentum behavior and power-law position-space tails leave characteristic imprints on the boundary response. In the second reading, $\lz$ is an AdS-invariant \emph{bulk proper length}, entering the bulk propagator through the same proper-time weighting. The two readings regulate different things. The boundary resolution breaks dilatations and softens boundary short-distance behavior. The bulk proper length preserves conformal symmetry exactly and instead softens bulk loops.

The meaning of ``T-duality-inspired'' is deliberately restricted. String T-duality exchanges momentum and winding sectors on compact directions and is not the same as holographic bulk/boundary duality. We borrow an effective form factor from zero-point-length models and impose it within an otherwise standard holographic setup. We neither perform a Buscher transformation nor derive a string compactification, and a smeared boundary source is not a bulk stress-energy tensor.

The boundary reading has a direct geometric counterpart. We show that the Bessel filter is exactly the bulk-to-boundary propagator of a scalar with $\nu=1$, evaluated at radial depth $z=\lz$. A measurement with resolution $\lz$ is therefore holographically a probe sourced from the surface $z=\lz$, in the spirit of smeared-operator constructions of bulk reconstruction \cite{HKLL}. The same surface provides a natural anchor for Ryu--Takayanagi surfaces \cite{RT,RTlong}, which yields finite entanglement entropies that retain the perimeter law and the universal sphere free energy \cite{CHM,JKPS}. The same anchoring also controls correlations between separated regions. For two intervals in $\mathrm{AdS}_3$ we obtain the mutual-information connectivity transition in closed form for arbitrary interval lengths, prove that a unique transition exists for arbitrary intervals and that equal intervals maximize the connectivity window at fixed total length, and show that subresolution intervals obey a universal scaling form in which the critical separation grows as $D_1D_2(D_1+D_2)/\lz^2$. Filtering external sources leaves the vacuum state unchanged, so this anchoring is an additional geometric prescription, and we keep the distinction between the two constructions explicit throughout.

The bulk reading raises a sharper question: can the prescribed propagator arise from a unitary holographic theory? A necessary condition is a K\"all\'en--Lehmann decomposition into AdS propagators with non-negative weights \cite{Penedones,PPTVV}. Related spectral methods constrain quantum field theories in curved spacetime \cite{LPSVS,MPS}. We derive the exact signed spectral decomposition of the zero-point-length propagator. It is AdS invariant and retains the original pole, but its continuum is negative immediately above threshold. At a fixed low spectral scale $M_{\max}$ the integrated signed correction is $O(\lz^2M_{\max}^2)$; this smallness does not restore positivity. The scale $L/\lz$ marks the boundary-dimension crossover where the modification becomes substantial, but neither a unitary effective theory nor a mandatory tower of new positive-norm operators follows from this result. We therefore treat the bulk prescription as a covariant Euclidean regulator and a diagnostic of the obstruction to a unitary completion.

The paper is organized as follows. In Sec.~\ref{sec:kernel} we review the proper-time origin of the zero-point-length form factor and derive its position-space kernel. Section~\ref{sec:bulk} sets up the bulk scalar in $\mathrm{AdS}_4$ and its holographically renormalized response. In Sec.~\ref{sec:sources} we impose the Bessel filter on the boundary source and obtain the exact bulk solution, the filtered two-point function and its coincidence limit, and in Sec.~\ref{sec:larger} we derive the logarithmic long-distance imprint of the filter. Section~\ref{sec:uv} discusses the sense in which $\lz$ acts as an intrinsic ultraviolet scale and shows that the present example is free of contact-term ambiguities. Section~\ref{sec:positivity} analyzes reflection positivity for spacetime and spatial smearing. In Sec.~\ref{sec:entropy} we anchor Ryu--Takayanagi surfaces at $z=\lz$ and compute the disk and interval entropies, and in Sec.~\ref{sec:mutualresolution} we extend the construction to two intervals and derive the mutual-information connectivity transition, the bound set by equal intervals, and the subresolution scaling form. Section~\ref{sec:extensions} treats massive fields and higher-point functions and distinguishes the construction from a radial cutoff. In Sec.~\ref{sec:holo} we establish the holographic origin of the filter, and in Sec.~\ref{sec:twofield} we describe a two-field auxiliary transfer prescription for the source map. Section~\ref{sec:bulkloops} turns to the bulk reading of $\lz$ and computes the finite one-loop tadpole and the resulting anomalous dimension. Section~\ref{sec:spectral} carries out the spectral test of holographic consistency. We conclude in Sec.~\ref{sec:conclusions}. All analytic results have, in addition, been checked by independent numerical evaluation.

\section{From zero-point length to a normalized source kernel}
\label{sec:kernel}

We begin with the flat-space Euclidean propagator of a massless scalar with zero-point length $\lz$ \cite{Padmanabhan,SSP}. In Schwinger proper time $s$, the ordinary propagator is $1/p^2=\int_0^\infty\dd s\,e^{-sp^2}$. The zero-point-length prescription multiplies the integrand by $e^{-\lz^2/(4s)}$, which leaves long proper times untouched but suppresses the region $s\lesssim\lz^2$, where the particle probes distances shorter than $\lz$. For positive Euclidean momentum magnitude $p$ the integral can be done in closed form:
\begin{equation}
 D_{\lz}(p)=\int_0^{\infty}\dd s\,
 e^{-sp^2-\lz^2/(4s)}
 =\frac{\lz}{p}K_1(\lz p)
 =\frac{\PP_{\lz}(p)}{p^2}.
 \label{eq:propagator}
\end{equation}
The second equality uses $\int_0^\infty\dd s\,s^{\mu-1}e^{-as-b/s}=2(b/a)^{\mu/2}K_\mu(2\sqrt{ab})$ with $\mu=1$, $a=p^2$ and $b=\lz^2/4$. The integrand is peaked at $s_*=\lz/(2p)$, so at large momentum the exponent at the saddle is $-\lz p$, and this is the origin of the exponential ultraviolet suppression. As $\lz\to0$ the weight $e^{-\lz^2/4s}$ tends to one and $D_{\lz}\to1/p^2$. The sign convention in Eq.~\eqref{eq:propagator} is the Euclidean one used in the effective models of Refs.~\cite{Nicolini,Gaete}.

The last form of Eq.~\eqref{eq:propagator} defines the dimensionless filter $\PP_{\lz}(p)=\lz p\,K_1(\lz p)$. Its role becomes clear once the propagator acts on a source. The field $\varphi=D_{\lz}j$ satisfies
\begin{equation}
 p^2\varphi(p)=\PP_{\lz}(p)j(p),
\end{equation}
which is the ordinary massless equation with the source $j$ replaced by $\PP_{\lz}j$. For a linear equation, then, modifying the propagator and filtering the source are the same operation. This is the observation that motivates the definition \eqref{eq:definition}. It is a statement about the effective field equation and does not derive the filter from a string worldsheet.

To see how the filter behaves, write $q=\lz p$. The small-argument series of $K_1$ gives
\begin{equation}
 \PP(q)=1+\frac{q^2}{2}
 \left[\ln\frac q2+\gamma_E-\frac12\right]
 +O(q^4\ln q),
 \label{eq:smallP}
\end{equation}
so at low momentum the filter reduces to one and long-wavelength physics is unaffected. The large-argument asymptotics of $K_1$ give instead
\begin{equation}
 \PP(q)=\sqrt{\frac{\pi q}{2}}\,e^{-q}
 \left[1+\frac{3}{8q}+O(q^{-2})\right],
 \label{eq:largeP}
\end{equation}
so at high momentum the filter decays exponentially on the scale $p\sim\lz^{-1}$. Between these limits the filter is monotonic: the identity $\dd[qK_1(q)]/\dd q=-qK_0(q)$ and the positivity of $K_0$ on the positive real axis imply that $\PP$ decreases from its value at the origin, and therefore
\begin{equation}
 0<\PP(q)\leq1,\qquad q\geq0.
\end{equation}
The filter never amplifies a mode; it only suppresses short wavelengths.

The logarithm in Eq.~\eqref{eq:smallP} has a structural consequence. A filter with a local derivative expansion would be a power series in $p^2$, of the form $1+c_1\lz^2p^2+c_2\lz^4p^4+\dots$. The term $q^2\ln q$ cannot be written this way, so the exact filter is not equivalent to any finite set of higher-derivative terms. This non-analyticity is what produces the heavy position-space tails discussed below and the logarithmic long-distance imprint of Sec.~\ref{sec:larger}.

We now turn to position space. The filter acts on sources by convolution, and it is useful to know the corresponding kernel explicitly. In $n$ flat Euclidean dimensions, with the convention $f(x)=\int\dd^np\,(2\pi)^{-n}e^{ip\cdot x}f(p)$, the inverse transform of $\PP_{\lz}$ follows from a second proper-time representation,
\begin{equation}
 \PP_{\lz}(p)=\frac{\lz^2}{4}
 \int_0^{\infty}\frac{\dd s}{s^2}
 e^{-sp^2-\lz^2/(4s)},
 \label{eq:Pproper}
\end{equation}
obtained from the same integral formula with $\mu=-1$. The momentum dependence now sits entirely in the Gaussian $e^{-sp^2}$, whose transform is $(4\pi s)^{-n/2}e^{-x^2/4s}$. The remaining proper-time integral, after the substitution $u=1/s$, is a Gamma function, and the result is
\begin{equation}
 \rho_{\lz}^{(n)}(x)=
 \frac{\Gamma[(n+2)/2]}{\pi^{n/2}}
 \frac{\lz^2}{(x^2+\lz^2)^{(n+2)/2}}.
 \label{eq:rho}
\end{equation}
Because the integral of a kernel over all of space equals its Fourier transform at zero momentum, the condition $\PP_{\lz}(0)=1$ translates into unit normalization,
\begin{equation}
 \int\dd^nx\,\rho_{\lz}^{(n)}(x)=1,
 \label{eq:rhonorm}
\end{equation}
which can also be checked by direct radial integration. The kernel is positive and smooth, and it has the scaling form $\rho_{\lz}^{(n)}(x)=\lz^{-n}\rho_1^{(n)}(x/\lz)$, so it tends to $\delta^{(n)}(x)$ as $\lz\to0$. Near the origin it is flat, with a core of width of order $\lz$. Section~\ref{sec:holo} shows that Eq.~\eqref{eq:rho} is also the normalized AdS bulk-to-boundary propagator of a field with $\nu=1$ at depth $z=\lz$, which explains this particular form.

The case relevant for the three-dimensional Euclidean boundary of Sec.~\ref{sec:sources} is $n=3$:
\begin{equation}
 \rho_{\lz}^{(3)}(r)=
 \frac{3\lz^2}{4\pi(r^2+\lz^2)^{5/2}}.
 \label{eq:rho3}
\end{equation}
The same profile has a familiar interpretation when the three coordinates are spatial. A static point mass $M$ smeared with $\rho_{\lz}^{(3)}$ has density $\rho_M(r)=M\rho_{\lz}^{(3)}(r)$, and the mass enclosed within a sphere of radius $r$ follows from an elementary radial integral,
\begin{equation}
 M(<r)=4\pi\int_0^r\dd r'\,r'^2\rho_M(r')=\frac{Mr^3}{(r^2+\lz^2)^{3/2}}.
 \label{eq:massprofile}
\end{equation}
For $r\ll\lz$ the density is constant, $\rho_M\simeq3M/(4\pi\lz^3)$, and $M(<r)$ grows as $r^3$. For $r\gg\lz$ the full mass is recovered, with $M(<r)\simeq M[1-3\lz^2/(2r^2)]$. This is the effective mass profile behind the regular metric of Ref.~\cite{Nicolini}. In our boundary problem, however, the three integration coordinates include Euclidean time obtained by continuation from a $(2+1)$-dimensional theory. There Eq.~\eqref{eq:rho3} is a spacetime source kernel and not a physical mass density. Smearing in space alone, for a $(2+1)$-dimensional boundary theory, uses $\rho_{\lz}^{(2)}$ instead, and we return to the difference between the two in Sec.~\ref{sec:positivity}.

The kernel has one further property that will matter below: it decays as a power, not exponentially. For $|x|\gg\lz$, Eq.~\eqref{eq:rho} gives
\begin{equation}
 \rho^{(n)}_{\lz}(x)\simeq\frac{\Gamma[(n+2)/2]}{\pi^{n/2}}\frac{\lz^2}{|x|^{n+2}} .
\end{equation}
This is fast enough for the kernel to be normalizable, but not fast enough for its second moment to exist. In radial coordinates,
\begin{equation}
 \int\dd^nx\,x^2\rho^{(n)}_{\lz}(x)
 =\frac{2\pi^{n/2}}{\Gamma(n/2)}\int_0^\infty\dd r\,r^{n+1}\rho^{(n)}_{\lz}(r)=\infty,
\end{equation}
since $r^{n+1}\rho^{(n)}_{\lz}(r)\sim1/r$, the integral diverges logarithmically at large $r$ in every dimension. This is the position-space counterpart of the term $\tfrac{q^2}{2}\ln q$ in Eq.~\eqref{eq:smallP}. For an isotropic kernel with finite second moment $\langle x^2\rangle$, expanding $e^{ip\cdot x}$ to second order gives $\PP=1-\langle x^2\rangle p^2/(2n)+\dots$, which is analytic at order $p^2$. The logarithm in the Bessel filter is therefore the momentum-space signature of the divergent second moment. Physically, $\lz$ sets the size of the core of the probe but does not define a variance. The probe keeps power-law tails, and these tails leave the logarithmic long-distance imprint derived in Sec.~\ref{sec:larger}.


\section{Bulk scalar and the ordinary holographic response}
\label{sec:bulk}

Before introducing any filter, we set up the standard holographic calculation for a free massless scalar in $\mathrm{AdS}_4$. This fixes our conventions, the counterterm, and the normalization of the boundary two-point function. The filtered response of Sec.~\ref{sec:sources} is built directly on top of the result \eqref{eq:ordinaryG} derived here.

We work in Euclidean $\mathrm{AdS}_4$ in Poincar\'e coordinates,
\begin{equation}
 \dd s^2=\frac{L^2}{z^2}(\dd z^2+\dd x_i\dd x_i),
 \qquad z>0,\quad i=1,2,3,
 \label{eq:metric}
\end{equation}
where $L$ is the AdS radius. The conformal boundary sits at $z=0$, and the coordinates $x_i$ become the coordinates of the three-dimensional Euclidean boundary theory. Small $z$ corresponds to short distances on the boundary, and large $z$ to long distances.

On this background we place a probe scalar with action
\begin{equation}
 \begin{split}
 I[\phi]&=\frac{\mathcal N}{2}\int\dd^4X\sqrt g\,
 g^{MN}\partial_M\phi\,\partial_N\phi\\
 &=\frac{\CC}{2}\int\dd z\,\dd^3x\,z^{-2}
 \left[(\partial_z\phi)^2+(\partial_i\phi)^2\right],
 \end{split}
 \label{eq:action}
\end{equation}
where $\CC\equiv\mathcal NL^2>0$. The second form follows from $\sqrt g=L^4/z^4$ and $g^{MN}=(z^2/L^2)\delta^{MN}$, which combine into the single factor $L^2/z^2$. For dimensionless $\phi$, $\mathcal N$ has mass dimension two and $\CC$ is dimensionless; in a holographic theory $\CC$ plays the role of the normalization of the boundary two-point function and typically scales with the number of degrees of freedom of the dual theory. We neglect backreaction on the metric and self-interactions of the scalar, so the problem is linear.

Translation invariance along the boundary lets us work mode by mode. Writing $\phi(z,x)=\intp e^{ip\cdot x}\phi(z,p)$, the field equation becomes an ordinary differential equation in $z$,
\begin{equation}
 \phi''-\frac2z\phi'-p^2\phi=0,
 \label{eq:eom}
\end{equation}
where primes denote $\partial_z$ and $p=|p|$. The substitution $\phi=z^{3/2}f(pz)$ turns Eq.~\eqref{eq:eom} into the modified Bessel equation of order $3/2$, whose solutions are elementary. The two independent solutions are $(1+pz)e^{-pz}$ and $(1-pz)e^{pz}$. For $p>0$ the second grows exponentially into the interior and is excluded by regularity at the Poincar\'e horizon $z\to\infty$. The regular solution with boundary value $j(p)$ is therefore
\begin{equation}
 \phi(z,p)=F(z,p)j(p),\qquad
 F(z,p)=(1+pz)e^{-pz}.
 \label{eq:bulkF}
\end{equation}
The transfer function $F$ equals one at the boundary and decays on the radial scale $z\sim p^{-1}$, so a boundary mode of wavelength $p^{-1}$ penetrates the bulk to a depth of the same order. The zero mode is the continuous $p\to0$ limit, $F=1$.

Expanding Eq.~\eqref{eq:bulkF} near the boundary gives
\begin{equation}
 \phi(z,p)=j(p)-\frac{p^2z^2}{2}j(p)
 +z^3a(p)+\cdots,\quad
 a(p)=\frac{p^3}{3}j(p).
 \label{eq:asymptotic}
\end{equation}
The terms $z^0$ and $z^3$ are the two independent falloffs allowed by Eq.~\eqref{eq:eom}: inserting $\phi\sim z^\Delta$ near $z=0$ gives $\Delta(\Delta-3)=0$. In standard quantization the coefficient of $z^0$ is the source $j$ and the coefficient of $z^3$ is related to the response, so the dual operator $\OO$ has dimension $\Delta=3$, in agreement with $m^2L^2=\Delta(\Delta-3)$ at $m=0$. The $z^2$ term is fixed locally by $j$ through the field equation and carries no new information. The coefficient $a(p)$, by contrast, is determined by regularity in the deep interior, and this is where the non-local information about the boundary state enters. Its momentum dependence $p^3$ is not a polynomial in $p^2$, which is why it produces a genuine non-local correlator below.

With the regular solution in hand, we turn to the on-shell action. It diverges because of the infinite volume near $z=0$, so we cut off the geometry at $z=\epsilon$ and remove the divergence with a local counterterm. On a solution of Eq.~\eqref{eq:eom}, an integration by parts in $z$ reduces the bulk integral to a boundary term. The outward normal at $z=\epsilon$ points toward decreasing $z$, which gives
\begin{equation}
 I_{\rm os}= -\frac{\CC}{2}\intp
 \epsilon^{-2}\phi(\epsilon,p)\,\partial_z\phi(\epsilon,-p).
\end{equation}
The contribution from $z\to\infty$ vanishes because the regular solution decays there. Using $\partial_zF=-p^2ze^{-pz}$ and expanding in $\epsilon$,
\begin{equation}
 I_{\rm os}=\frac{\CC}{2}\intp j(p)j(-p)
 \left[\frac{p^2}{\epsilon}-p^3+O(\epsilon)\right].
\end{equation}
The divergent term is proportional to $p^2$, which in position space is $(\partial_i j)^2$. It is therefore local in the source, and it can be cancelled by a covariant local functional of the field on the cutoff surface. The appropriate counterterm is
\begin{equation}
 I_{\rm ct}=-\frac{\mathcal NL}{2}
 \int_{z=\epsilon}\dd^3x\sqrt\gamma\,
 \gamma^{ij}\partial_i\phi\,\partial_j\phi
 =-\frac{\CC}{2\epsilon}\int\dd^3x\,(\partial_i\phi)^2,
\end{equation}
where $\gamma_{ij}=(L^2/\epsilon^2)\delta_{ij}$ is the induced metric, so that $\sqrt\gamma\,\gamma^{ij}=(L/\epsilon)\delta_{ij}$. Since $\phi(\epsilon,p)=j(p)+O(\epsilon^2)$, this term removes exactly the $1/\epsilon$ divergence and leaves the finite part untouched in the limit $\epsilon\to0$.

Holographic renormalization allows one further ambiguity: finite local counterterms quadratic in the source could be added with arbitrary coefficients. Setting all of them to zero defines the \emph{minimal scheme} used below. The renormalized action and generating functional are then
\begin{equation}
 I_{\rm ren}[j]=-\frac{\CC}{2}\intp p^3j(p)j(-p),
 \qquad W_0[j]=-I_{\rm ren}[j].
 \label{eq:ren}
\end{equation}
The relation $Z_0[j]\simeq\exp(-I_{\rm ren}[j])$ is the classical bulk approximation to the generating functional of the boundary theory, valid at leading order in the large-$\CC$ expansion. For the present $\Delta=3$ operator the minimal scheme is in fact the only scale-covariant choice, as we show in Sec.~\ref{sec:uv}. The one-point function follows by varying $W_0$ with respect to the source, $\delta W_0/\delta j(-p)=\CC p^3j(p)=3\CC\,a(p)$, so the response coefficient $a(p)$ of Eq.~\eqref{eq:asymptotic} is the expectation value of $\OO$ up to the factor $3\CC$.

A second variation gives the connected two-point function. Writing $\langle\OO(p)\OO(q)\rangle_c=(2\pi)^3\delta^{(3)}(p+q)\mathcal G_0(p)$, we find
\begin{equation}
 \mathcal G_0(p)=\CC p^3,\qquad
 \mathcal G_0(x)=\frac{12\CC}{\pi^2|x|^6}
 \quad (x\ne0).
 \label{eq:ordinaryG}
\end{equation}
The position-space form follows from the Fourier transform of $p^3$, and it is the power law fixed by conformal invariance for an operator of dimension three, with $\CC$ setting its normalization. The momentum-space expression is the more complete of the two. The function $|x|^{-6}$ is not integrable at the origin in three dimensions, so on its own it does not define a distribution; it has to be extended to $x=0$, and the extension is what $\CC p^3$ provides. This point is irrelevant for correlators at separated points but becomes essential once the operator is smeared, because smearing integrates over the coincident region. For this reason the filtered correlators of Sec.~\ref{sec:sources} are always constructed from the momentum-space form \eqref{eq:ordinaryG} rather than by convolving $|x|^{-6}$ directly.

\section{T-duality-inspired boundary sources}
\label{sec:sources}

We now combine the two ingredients of the previous sections. The zero-point-length filter of Sec.~\ref{sec:kernel} is imposed on the source of the holographic scalar of Sec.~\ref{sec:bulk}, and we follow its effect through the bulk solution, the generating functional and the boundary two-point function.

We prescribe the asymptotic value of the bulk field to be the filtered version of an independent source $J$:
\begin{equation}
 j(p)=\PP_{\lz}(p)J(p),\qquad
 j(x)=\int\dd^3y\,\rho_{\lz}^{(3)}(x-y)J(y).
 \label{eq:sourcemap}
\end{equation}
The two forms are equivalent, since multiplication by $\PP_{\lz}$ in momentum space is convolution with $\rho^{(3)}_{\lz}$ in position space. The roles of the two sources should be kept apart. The function $J$ is the source an experimenter controls; the function $j$ is the Dirichlet datum that the bulk field actually sees at $z=0$. Nothing else in the holographic setup is changed.

Because the map \eqref{eq:sourcemap} is linear, the filter can be moved from the source onto the operator. Inserting $j=\rho^{(3)}_{\lz}*J$ into the source coupling $\int\dd^3x\,j\,\OO$ and using that the kernel is real and even, $\rho^{(3)}_{\lz}(x-y)=\rho^{(3)}_{\lz}(y-x)$, gives the exact identity
\begin{equation}
 \frac{Z_{\lz}[J]}{Z_0[0]}
 \equiv\frac{Z_0[\PP_{\lz}J]}{Z_0[0]}
 =\left\langle e^{\int\dd^3x\,J(x)\OO_{\lz}(x)}\right\rangle_0,
 \label{eq:Zidentity}
\end{equation}
in which the source couples to the smeared operator
\begin{equation}
 \OO_{\lz}(x)=\int\dd^3y\,
 \rho_{\lz}^{(3)}(x-y)\OO(y).
 \label{eq:operator}
\end{equation}
The expectation value in Eq.~\eqref{eq:Zidentity} is taken in the original theory. Smearing the source is therefore the same as smearing the operator, and neither changes the source-free Hamiltonian or the vacuum. What changes is the observable: $J$ couples to $\OO_{\lz}$, a local operator averaged over a region of size $\lz$. At fixed $\lz$ the smeared operator is not a local conformal primary, even though $\OO$ is one, because $\lz$ is a fixed length and conformal transformations would rescale it.

We next determine the bulk field. The bulk equation is linear and its regular solution \eqref{eq:bulkF} holds for any boundary value, so the filtered solution follows by substituting $j=\PP_{\lz}J$:
\begin{equation}
 \phi_{\lz}(z,p)=
 (1+pz)e^{-pz}\,[\lz pK_1(\lz p)]\,J(p).
 \label{eq:mainbulk}
\end{equation}
This expression is exact. The field equation, the mass and the radial profile $(1+pz)e^{-pz}$ are those of the ordinary problem; only the amplitude with which each boundary mode enters has been modified.

In position space the solution takes the form
\begin{equation}
 \phi_{\lz}(z,x)=\int\dd^3y\,K_{\lz}(z,x-y)J(y),
\end{equation}
where the effective kernel is the ordinary bulk-to-boundary propagator convolved with the source kernel,
\begin{equation}
 K_{\lz}(z,x)=\int\dd^3u\,K_0(z,x-u)\rho_{\lz}^{(3)}(u).
\end{equation}
Here $K_0$ is the $\Delta=3$ bulk-to-boundary propagator,
\begin{equation}
 K_0(z,x)=\frac4{\pi^2}\frac{z^3}{(z^2+x^2)^3},
\end{equation}
normalized so that $\int\dd^3x\,K_0(z,x)=1$ at every $z$. The smeared kernel $K_{\lz}$ is an average of $\rho^{(3)}_{\lz}$ with a normalized positive weight, so it cannot exceed the maximum of $\rho^{(3)}_{\lz}$:
\begin{equation}
 K_{\lz}(z,x)\leq\rho_{\lz}^{(3)}(0)=\frac{3}{4\pi\lz^3}.
\end{equation}
For a point-like source $J=\delta^{(3)}(x)$ the ordinary field diverges at the insertion as $z\to0$, while the filtered field approaches the bounded profile $\rho^{(3)}_{\lz}(x)$. This is a statement about the amplitude of the probe scalar. The background metric is untouched, and nothing is implied about spacetime curvature.

The generating functional follows by composing the renormalized action \eqref{eq:ren} with the source map. Each factor of $j$ brings one factor of the filter:
\begin{equation}
 W_{\lz}[J]=\frac{\CC}{2}\intp
 p^3\PP_{\lz}(p)^2J(p)J(-p).
\end{equation}
Two functional derivatives with respect to $J$, evaluated at $J=0$, give the filtered two-point function,
\begin{equation}
 \mathcal G_{\lz}(p)
 =\CC p^3[\lz pK_1(\lz p)]^2.
 \label{eq:mainG}
\end{equation}
The filter appears squared because both operator insertions are smeared: $\langle\OO_{\lz}\OO_{\lz}\rangle$ carries one factor of $\PP_{\lz}$ per insertion. The same counting appears in the one-point function. Let $a_{\lz}=p^3\PP_{\lz}J/3$ be the coefficient of $z^3$ in Eq.~\eqref{eq:mainbulk}. The measured response is
\begin{equation}
 \frac{\delta W_{\lz}}{\delta J(-p)}
 =3\CC\,\PP_{\lz}(p)\,a_{\lz}(p).
\end{equation}
The bulk coefficient $a_{\lz}$ already contains one filter from the source. The variation with respect to $J$, rather than $j$, adds a second one, because the experimenter measures the smeared operator $\OO_{\lz}$ and not $\OO$. Reading the response as $3\CC a_{\lz}$ would correspond to an unsmeared operator in the presence of a smeared source.

For the same reason, Eq.~\eqref{eq:mainG} should not be identified with the flat-space propagator \eqref{eq:propagator}. There only the source of a linear equation was filtered, and a single factor of $\PP_{\lz}$ appeared. A holographic prescription with a single factor would require a different probe. Two examples are the symmetric map $j=\PP_{\lz}^{1/2}J$ and a mixed correlator with one smeared and one unsmeared insertion.

The effect of the filter is cleanest in the ratio of the filtered and ordinary responses,
\begin{equation}
 R(q)\equiv\frac{\mathcal G_{\lz}(p)}{\mathcal G_0(p)}
 =[qK_1(q)]^2,\qquad q=\lz p,
\end{equation}
which is independent of the normalization $\CC$ and depends on momentum only through $q$. Squaring the small-$q$ expansion \eqref{eq:smallP} gives
\begin{equation}
 R(q)=1+q^2\left[\ln\frac q2+\gamma_E-\frac12\right]
 +O(q^4\ln^2q),
 \label{eq:Rsmall}
\end{equation}
so modes with $p\ll\lz^{-1}$ are measured almost exactly, with a relative correction of order $q^2\ln q$. Squaring the large-$q$ expansion \eqref{eq:largeP} gives instead
\begin{equation}
 R(q)=\frac{\pi q}{2}e^{-2q}
 \left[1+\frac{3}{4q}+O(q^{-2})\right],
\end{equation}
so modes with $p\gg\lz^{-1}$ are suppressed exponentially. The leading large-momentum response is therefore
\begin{equation}
 \mathcal G_{\lz}(p)\sim
 \frac{\pi\CC\lz}{2}\,p^4e^{-2\lz p}.
\end{equation}
The rate $2\lz$ in the exponent is the sum of the suppressions from the two insertions.

\begin{figure*}[t]
 \centering
 \includegraphics[width=\textwidth]{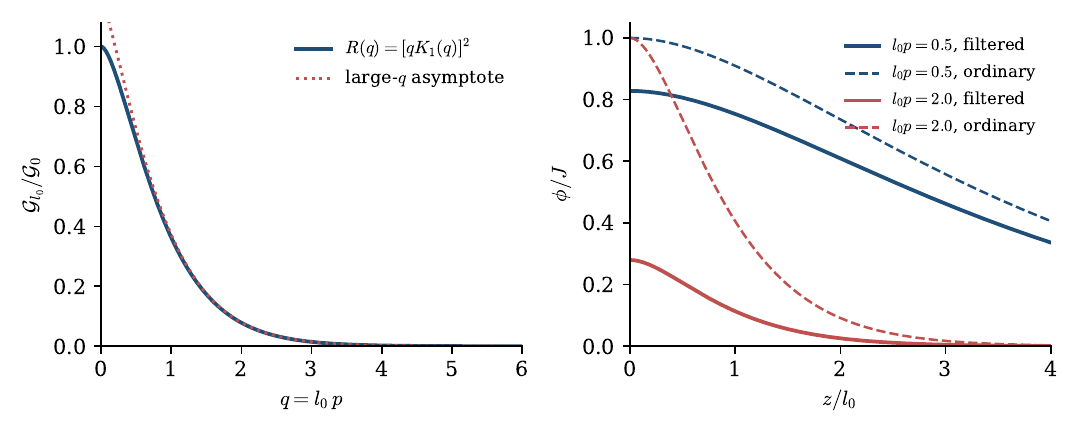}
 \caption{Left: exact suppression of the Euclidean scalar response, $R(q)=[qK_1(q)]^2$, with its large-$q$ asymptote. Right: bulk transfer amplitudes $\phi/J$ versus $z/\lz$ for two momenta, with filtered (solid) and ordinary (dashed) sources. Filtering changes the boundary amplitude at fixed nominal source $J$; the radial equation and its regular solution are unchanged.}
 \label{fig:response}
\end{figure*}

The filter does not change the scaling dimension of the underlying operator. The radial powers $z^0$ and $z^3$ in the near-boundary expansion are unchanged, so $\OO$ keeps $\Delta=3$; the length $\lz$ characterizes the measurement and produces no anomalous dimension. The ordinary CFT correlator is recovered at fixed $p$ as $\lz\to0$. The limit is not uniform in $p$, since for any $\lz>0$ there are momenta $p\gtrsim\lz^{-1}$ at which the filtered and unfiltered responses differ by an arbitrarily large factor. Figure~\ref{fig:response} shows the suppression ratio and the corresponding bulk profiles.

In position space, the rotationally symmetric three-dimensional Fourier transform turns Eq.~\eqref{eq:mainG} into a one-dimensional integral. In the minimal scheme,
\begin{equation}
 \mathcal G_{\lz}(r)=\frac{\CC}{2\pi^2}
 \int_0^{\infty}\dd p\,p^5
 [\lz pK_1(\lz p)]^2\frac{\sin(pr)}{pr},
 \label{eq:realG}
\end{equation}
with $r=|x|$. Without the filter this integral would not converge, which is the momentum-space form of the non-integrable singularity of $|x|^{-6}$. With the filter the integrand decays as $p^6e^{-2\lz p}$ and the integral converges absolutely for every $r$, including $r=0$. At coincidence $\sin(pr)/(pr)\to1$, and the substitution $q=\lz p$ reduces the integral to a Bessel moment. Such moments follow from the standard formula
\begin{equation}
 \begin{split}
 \int_0^{\infty}\dd q\,q^{\mu-1}K_{\nu}(q)^2
 ={}&\frac{2^{\mu-3}}{\Gamma(\mu)}
 \Gamma\!\left(\frac\mu2\right)^2\\
 &\times\Gamma\!\left(\frac\mu2+\nu\right)
 \Gamma\!\left(\frac\mu2-\nu\right),
 \end{split}
 \label{eq:Besselmoment}
\end{equation}
valid for $\operatorname{Re}\mu>2|\operatorname{Re}\nu|$. For $\mu=8$ and $\nu=1$ it gives $\int_0^{\infty}\dd q\,q^7K_1(q)^2=384/35$, and the coincidence value is therefore
\begin{equation}
 \mathcal G_{\lz}(0)
 =\frac{192\,\CC}{35\pi^2\lz^6}.
 \label{eq:coincidence}
\end{equation}
The scaling $\lz^{-6}$ is what dimensional analysis requires for an operator of dimension three measured with resolution $\lz$. Since every momentum moment converges, $\mathcal G_{\lz}(r)$ is smooth at $r=0$ and can be expanded in powers of $r^2$ there. Equation~\eqref{eq:coincidence} is a finite Euclidean response in a specified scheme. It should not automatically be read as the equal-time variance of a time-local Lorentzian observable, for reasons explained in Sec.~\ref{sec:positivity}.

\section{Long-distance behavior: the logarithmic imprint of \texorpdfstring{$\lz$}{l0}}
\label{sec:larger}
The filtered correlator \eqref{eq:realG} is finite at coincidence. At large separations, on the other hand, one expects the filter to become invisible, and the question is how fast. The answer is controlled by the small-momentum behavior of $\mathcal G_{\lz}(p)$. Terms analytic in $p^2$, such as $p^{2k}$, Fourier transform to derivatives of $\delta^{(3)}(x)$ and do not contribute at $r\neq0$. The large-$r$ expansion therefore comes entirely from the terms of $\mathcal G_{\lz}(p)$ that are non-analytic at $p=0$. We derive the leading correction in five steps.

\emph{Step 1: small-momentum expansion.} Multiplying $\mathcal G_0=\CC p^3$ by the small-$q$ form \eqref{eq:Rsmall} of $R(q)$ gives
\begin{multline}
 \mathcal G_{\lz}(p)=\CC p^3+\CC\lz^2p^5\left[\ln p+\ln\frac{\lz}{2}+\gamma_E-\frac12\right]\\
 +O\!\left(\lz^4p^7\ln^2p\right).
 \label{eq:Gsmallp}
\end{multline}
Both $p^5$ and $p^5\ln p$ are non-analytic, so both contribute at separated points.

\emph{Step 2: Fourier transform of a power.} By analytic continuation in $\nu$, for $x\neq0$,
\begin{equation}
 \int\frac{\dd^3p}{(2\pi)^3}e^{ip\cdot x}\,p^{2\nu}=\frac{c(\nu)}{|x|^{3+2\nu}},
 \qquad
 c(\nu)=\frac{4^\nu\,\Gamma\!\left(\tfrac32+\nu\right)}{\pi^{3/2}\,\Gamma(-\nu)} .
 \label{eq:FTpower}
\end{equation}
The factor $1/\Gamma(-\nu)$ vanishes at non-negative integers $\nu$, which is how this formula encodes the fact that analytic terms give only contact contributions. For $\nu=3/2$, $\Gamma(-3/2)=4\sqrt\pi/3$ gives $c(3/2)=12/\pi^2$, which reproduces Eq.~\eqref{eq:ordinaryG}. For $\nu=5/2$, $\Gamma(4)=6$ and $\Gamma(-5/2)=-8\sqrt\pi/15$ give
\begin{equation}
 c(5/2)=\frac{32\cdot6}{\pi^{3/2}\,(-8\sqrt\pi/15)}=-\frac{360}{\pi^2}.
\end{equation}

\emph{Step 3: the logarithm.} A logarithm is generated by differentiating a power with respect to its exponent, $p^{2\nu}\ln p=\tfrac12\partial_\nu p^{2\nu}$. Applying this to Eq.~\eqref{eq:FTpower},
\begin{equation}
 \begin{split}
 \int\frac{\dd^3p}{(2\pi)^3}e^{ip\cdot x}\,p^{2\nu}\ln p
 &=\frac12\partial_\nu\!\left[\frac{c(\nu)}{r^{3+2\nu}}\right]\\
 &=\frac{c(\nu)}{r^{3+2\nu}}\left[\frac12\,\partial_\nu\ln c(\nu)-\ln r\right].
 \end{split}
\end{equation}
A logarithm of momentum thus becomes a logarithm of distance, with the opposite sign.

\emph{Step 4: the digamma functions.} From Eq.~\eqref{eq:FTpower}, $\partial_\nu\ln c=\ln4+\psi(\tfrac32+\nu)+\psi(-\nu)$, where $\psi$ is the digamma function. At $\nu=5/2$ we need $\psi(4)=\tfrac{11}{6}-\gamma_E$ and $\psi(-\tfrac52)$. The latter follows from $\psi(\tfrac12)=-\gamma_E-2\ln2$ by applying the recurrence $\psi(x+1)=\psi(x)+1/x$ three times downward:
\begin{equation}
 \psi\!\left(-\tfrac52\right)=\psi\!\left(\tfrac12\right)+2+\tfrac23+\tfrac25=-\gamma_E-2\ln2+\tfrac{46}{15}.
\end{equation}
Adding $\ln4$ and $\psi(4)$, the $\ln2$ terms cancel and we obtain
\begin{equation}
 \frac12\,\partial_\nu\ln c\Big|_{\nu=5/2}=\frac{147}{60}-\gamma_E .
\end{equation}

\emph{Step 5: assembly.} Transforming the $p^5$ and $p^5\ln p$ terms of Eq.~\eqref{eq:Gsmallp} with Steps 2 to 4, the two Euler constants cancel:
\begin{equation}
 \begin{split}
 &\CC\lz^2\,\frac{c(5/2)}{r^8}\left[\frac{147}{60}-\gamma_E-\ln r+\ln\frac{\lz}{2}+\gamma_E-\frac12\right]\\
 &\qquad=\frac{360\,\CC\lz^2}{\pi^2r^8}\left[\ln\frac{2r}{\lz}-\frac{39}{20}\right].
 \end{split}
\end{equation}
Dividing by $\mathcal G_0(r)=12\CC/(\pi^2r^6)$ and adding the leading conformal term gives the large-distance expansion
\begin{multline}
 \mathcal G_{\lz}(r)=\frac{12\,\CC}{\pi^2r^6}\left[1+30\,\frac{\lz^2}{r^2}\left(\ln\frac{2r}{\lz}-\frac{39}{20}\right)\right.\\
 \left.+O\!\left(\frac{\lz^4}{r^4}\ln^2\frac{r}{\lz}\right)\right].
 \label{eq:largeR}
\end{multline}

The correction decays as $(\lz/r)^2$, as expected, but it carries a logarithm and is positive for $r>\tfrac12e^{39/20}\lz\approx3.5\lz$. The logarithm characterizes the nonanalytic low-momentum structure of this zero-point-length filter; it is not by itself a unique signature of T-duality. A smearing kernel with a finite second moment would give an analytic $p^2$ correction to $\PP_{\lz}$, hence a correction to $\mathcal G_{\lz}$ proportional to $p^5$ without a logarithm, and a pure $(\lz/r)^2$ term in position space. The $\ln(r/\lz)$ is the long-distance trace of the heavy tails of $\rho^{(n)}_{\lz}$ found in Sec.~\ref{sec:kernel}. Equation~\eqref{eq:largeR} is fixed within the scale-covariant parent prescription of Sec.~\ref{sec:uv}. An arbitrary additional contact convention, when subsequently smeared, can also affect separated points through the tails of the filter.

\section{An intrinsic ultraviolet scale without a hard momentum cutoff}
\label{sec:uv}
The calculation above uses no momentum cutoff. The length $\lz$ supplies a soft ultraviolet scale $\Lambda_{\rm eff}\sim\lz^{-1}$, and the momentum integral in Eq.~\eqref{eq:realG} runs over the full range $0\leq p<\infty$ rather than $p\leq\Lambda_{\rm UV}$. The same holds for every momentum moment of the response. For non-negative integer $n$, the substitution $q=\lz p$ gives
\begin{equation}
 \int\frac{\dd^3p}{(2\pi)^3}\,p^n\mathcal G_{\lz}(p)
 =\frac{\CC}{2\pi^2\lz^{n+6}}
 \int_0^{\infty}\dd q\,q^{n+7}K_1(q)^2<\infty,
 \label{eq:moments}
\end{equation}
which is finite because $K_1(q)^2$ decays as $e^{-2q}$. Without the filter all of these integrals diverge at large $p$. The divergence of the unfiltered response is replaced by a finite value set by the source resolution, and there is no extra cutoff parameter to send to infinity at the end. Modes with $p\gg\lz^{-1}$ are exponentially suppressed rather than excluded, so $\lz$ behaves as a soft resolution length, not as a sharp forbidden distance or a lattice spacing.

This ultraviolet scale has to be kept separate from the radial divergence of the AdS action, which the filter does not remove. Repeating the calculation of Sec.~\ref{sec:bulk} with $j=\PP_{\lz}J$, the on-shell action before the counterterm is added is
\begin{equation}
 \begin{split}
 I_{{\rm os},\lz}[J]
 &=\frac{\CC}{2\epsilon}\intp
 p^2\PP_{\lz}(p)^2J(p)J(-p)\\
 &\quad-\frac{\CC}{2}\intp p^3\PP_{\lz}(p)^2J(p)J(-p)
 +O(\epsilon).
 \end{split}
 \label{eq:remainingdiv}
\end{equation}
For reasonable sources the momentum integral in the first term is finite at fixed $\lz>0$, but it is multiplied by $1/\epsilon$, which still diverges as the cutoff surface approaches the boundary. Source smearing cannot cancel this factor, because it comes from the volume of AdS near $z=0$ rather than from large momenta. We therefore keep the usual local counterterm, written in terms of $j$, remove $\epsilon$, and only then evaluate the remaining momentum integrals without a cutoff. The auxiliary radial cutoff $\epsilon$ is not identified with $\lz$.

The precise statement is therefore the following. For the filtered quadratic Euclidean response and its moments, $\lz$ replaces an independent hard momentum cutoff. It does not replace holographic renormalization, and filtering external sources does not regulate quantum loops, either in the bulk or in the unsmeared boundary theory.

We now return to the scheme dependence mentioned in Sec.~\ref{sec:bulk}. Holographic renormalization allows finite local counterterms quadratic in the source, and in general these change the two-point function by a polynomial in $p^2$. Suppose another admissible scheme gives $\mathcal G_0(p)\mapsto\CC p^3+Q(p^2)$ with $Q$ a polynomial. Composing with the source map then gives
\begin{equation}
 \mathcal G_{\lz}(p)\longmapsto
 \PP_{\lz}(p)^2\left[\CC p^3+Q(p^2)\right].
 \label{eq:contact}
\end{equation}
Before smearing, $Q$ contributes only contact terms supported at $x=0$. After smearing, $\PP_{\lz}^2Q$ is no longer a polynomial, and its Fourier transform is a function of width of order $\lz$ that also affects nearby separated points. In a theory where such a $Q$ is allowed, Eq.~\eqref{eq:coincidence} would therefore be scheme dependent, and Eqs.~\eqref{eq:mainG} and \eqref{eq:realG} would refer to the scheme \eqref{eq:ren}.

The same consideration explains why the construction cannot be carried out by convolving the separated-point function $|x|^{-6}$ directly. That function must first be extended to a distribution at $x=0$, and the extension is the choice of contact terms. Our procedure is to renormalize $W_0$, fix its contact terms, and only then compose with $\PP_{\lz}$. Note also that local counterterms in the Dirichlet datum $j$ need not be local when written in terms of the nominal source $J$. Whatever the finite polynomial $Q$, exponential suppression at large momentum and finiteness at coincidence survive, since both come from the filter; only the finite coefficients depend on the scheme. No conclusion about loop finiteness, finite stress-energy or regularity of a backreacted metric follows from this quadratic probe calculation.

The absence of an ambiguity for the $\Delta=3$ probe requires a qualification: we restrict the parent quadratic functional to a scale-covariant renormalization prescription on a flat boundary. Finite local counterterms with dimensionful coefficients can otherwise be added also for noninteger $\nu$; integer $\nu$ specifically marks the logarithmic resonance and its associated scale dependence. Within the scale-covariant prescription, the $\Delta=3$ response is fixed for three related reasons.
\begin{enumerate}[leftmargin=*,itemsep=2pt]
 \item For $d=3$ and $\Delta=3$ the Dirichlet datum $j$ is dimensionless. A finite local counterterm $\int\dd^3x\,j\,(-\partial^2)^n j$ therefore contributes $Q(p^2)=a_np^{2n}$ with a coefficient of mass dimension $[a_n]=3-2n$.
 \item The parent CFT is renormalized before the filter is applied. Requiring its quadratic source functional to remain scale covariant excludes dimensionful contact coefficients. A nonzero scale-covariant coefficient would require $3-2n=0$, which has no integer solution, so $Q\equiv0$ within this prescription. Allowing an arbitrary subtraction scale would permit additional contact conventions, at the expense of this requirement.
 \item The same conclusion follows from the bulk. Since $\nu=3/2$ is half-integer, the near-boundary expansion contains no $z^{2\nu}\ln z$ term and requires no $\ln\epsilon$ counterterm. In position space, $|x|^{-2\Delta}$ on $\mathbb R^3$ has poles in $\Delta$ only at $2\Delta=3+2k$ with $k=0,1,2,\dots$. For $\Delta=3$ the combination $2\Delta-d=3$ is odd, so $|x|^{-6}$ has a unique scale-covariant extension to a distribution.
\end{enumerate}
Consequently Eqs.~\eqref{eq:mainG}, \eqref{eq:realG}, \eqref{eq:coincidence} and \eqref{eq:largeR} are fixed up to $\CC$ within the scale-covariant parent prescription. This is not independence from arbitrary finite local counterterms. Integer $\nu$, as in the $\Delta=5/2$ channel of Sec.~\ref{sec:twofield}, additionally entails logarithmic scale dependence.

\section{Positivity: spacetime versus spatial smearing}
\label{sec:positivity}

The full Euclidean filter depends on $p=\sqrt{\omega_E^2+\bm k^2}$, so it smears in Euclidean time $\tau$ as well as in space. Smearing in time is non-local in $\tau$, and this raises a question that does not arise for spatial smearing: whether the filtered correlator can still be the two-point function of an operator in a unitary theory with a Hamiltonian. The relevant criterion is Osterwalder--Schrader reflection positivity \cite{OS}. Positivity of $\mathcal G_{\lz}(p)$ as a function of Euclidean momentum does not imply it, and in fact Eq.~\eqref{eq:mainG}, read as the covariance of a time-local operator, violates it.

To show this, fix the spatial momentum $\bm k$ and transform to Euclidean time:
\begin{equation}
 C_{\lz}(\tau,\bm k)=\int\frac{\dd\omega_E}{2\pi}
 e^{i\omega_E\tau}\mathcal G_{\lz}
 \!\left(\sqrt{\omega_E^2+\bm k^2}\right).
\end{equation}
The exponential damping of $\mathcal G_{\lz}$ at large $\omega_E$ makes $C_{\lz}$ smooth, and it is even in $\tau$ because $\mathcal G_{\lz}$ depends on $\omega_E^2$. Differentiating twice under the integral sign at $\tau=0$ brings down $-\omega_E^2$, and since $\mathcal G_{\lz}>0$,
\begin{equation}
 C_{\lz}''(0,\bm k)=
 -\int\frac{\dd\omega_E}{2\pi}\,\omega_E^2\,
 \mathcal G_{\lz}\!\left(\sqrt{\omega_E^2+\bm k^2}\right)<0.
 \label{eq:RPfail}
\end{equation}
By continuity, $C_{\lz}''(\tau,\bm k)<0$ also for small $\tau>0$: the correlator is concave in Euclidean time near the origin.

A reflection-positive, translation-invariant two-point function cannot behave this way. For $\tau>0$ it admits the spectral representation
\begin{equation}
 C(\tau,\bm k)=\int_0^{\infty}\dd E\,
 \sigma(E,\bm k)e^{-E\tau},\qquad \sigma\geq0,
\end{equation}
which expresses $C$ as a positive sum of decaying exponentials, one for each energy eigenstate. Differentiating twice gives $C''=\int\dd E\,E^2\sigma\,e^{-E\tau}\geq0$, so such a function is convex in $\tau$, in contradiction with Eq.~\eqref{eq:RPfail}.

The result does not contradict unitarity of the parent CFT. Reflection positivity applies to operators supported at positive Euclidean time, and an insertion smeared in time with the kernel $\rho^{(3)}_{\lz}$ and centered at $\tau>0$ has tails extending to negative $\tau$. The Euclidean source functional remains a well-defined filtered probe of a unitary theory, but it does not define a new unitary local field theory.

Reflection positivity can, however, be kept by smearing only in space. In a Lorentzian $(2+1)$-dimensional boundary theory we define
\begin{equation}
 \OO_{\lz}(t,\bm x)=\int\dd^2y\,
 \rho_{\lz}^{(2)}(\bm x-\bm y)\OO(t,\bm y),
\end{equation}
with the two-dimensional kernel
\begin{equation}
 \rho_{\lz}^{(2)}(\bm x)=
 \frac{\lz^2}{\pi(\bm x^2+\lz^2)^2},
\end{equation}
which is the $n=2$ case of Eq.~\eqref{eq:rho}. The operator is local in time, so it acts within the Hilbert space at a fixed instant. In the same parent scheme, with $k=|\bm k|$, its Euclidean two-point function is
\begin{equation}
 \mathcal G_{\lz,E}(\omega_E,\bm k)
 =\PP_{\lz}(k)^2\,\CC(\omega_E^2+k^2)^{3/2}.
\end{equation}
The filter now depends only on the spatial momentum. With the standard infalling condition in the bulk interior, the retarded response is obtained in the same way:
\begin{equation}
 \mathcal G_{\lz,R}(\omega,\bm k)
 =\PP_{\lz}(k)^2\,\mathcal G_{0,R}(\omega,\bm k).
 \label{eq:retarded}
\end{equation}
No analytic continuation of a frequency-dependent Bessel filter is needed. At each $\bm k$, the positive Wightman spectral measure of the parent theory is multiplied by the non-negative number $\PP_{\lz}(k)^2$, so the norm of every state created by the smeared operator stays positive. In Euclidean language, spatial smearing preserves the support of an operator on each side of a time-reflection plane.

The price is a preferred time slicing. Spatial smearing is not boost invariant. The smeared operator is non-local in space, so we do not claim microcausality for it. High frequencies at fixed $k$ are not suppressed, and as a result spatial smearing does not make the full spacetime coincidence limit finite for this $\Delta=3$ operator. Equations~\eqref{eq:coincidence} and \eqref{eq:RPfail} refer to the fully Euclidean prescription only. Table~\ref{tab:compare} lists the properties each construction retains.

\begin{table*}[t]
\centering
\begin{tabular}{lcc}
\hline
Property & Spacetime smearing $\rho^{(3)}_{\lz}$ & Spatial smearing $\rho^{(2)}_{\lz}$\\
\hline
Euclidean $O(3)$ invariance & yes & no\\
Exponential suppression at large $p$ & all directions & spatial $k$ only\\
Finite coincidence limit & yes, Eq.~\eqref{eq:coincidence} & no\\
Reflection positivity (time-local reading) & violated, Eq.~\eqref{eq:RPfail} & preserved\\
Retarded response & not specified as a unitary response & $\PP_{\lz}(k)^2\mathcal G_{0,R}$\\
\hline
\end{tabular}
\caption{Comparison of the two finite-resolution prescriptions.}
\label{tab:compare}
\end{table*}

\section{Entanglement entropy at finite resolution}
\label{sec:entropy}

The scalar probe alone does not determine an entanglement entropy, since the Ryu--Takayanagi (RT) formula is a property of the gravitational sector. For this section we assume a semiclassical Einstein sector,
\begin{equation}
 I_{\rm grav}=\frac{1}{16\pi G_4}\int\dd^4X\sqrt{-g}
 \left(\mathcal R+\frac6{L^2}\right)+I_{\rm bdy},
\end{equation}
where $I_{\rm bdy}$ contains the usual boundary terms. At $J=0$ and leading classical order the solution is the Lorentzian AdS vacuum, which we write in polar coordinates on the boundary:
\begin{equation}
 \dd s^2=\frac{L^2}{z^2}
 \left(-\dd t^2+\dd z^2+\dd r^2+r^2\dd\theta^2\right).
 \label{eq:RTmetric}
\end{equation}
The scalar normalization $\CC$ and the gravitational normalization $L^2/G_4$ are independent parameters, and the second one sets the scale of all entropies below.

For a spatial region $A$ on a constant-time slice, the RT prescription is \cite{RT,RTlong}
\begin{equation}
 S_A=\frac{\operatorname{Area}(\gamma_A)}{4G_4},
 \qquad \partial\gamma_A=\partial A,\qquad \gamma_A\sim A,
 \label{eq:RT}
\end{equation}
where $\gamma_A$ is the bulk surface of minimal area that ends on the boundary of $A$ and is homologous to $A$. When $\gamma_A$ is anchored on the conformal boundary $z=0$, its area diverges because of the warp factor $L^2/z^2$; this is the holographic form of the ultraviolet divergence of entanglement entropy in quantum field theory. To model finite resolution we move the anchor to the surface $z=\lz$ and define
\begin{equation}
 S_A^{\rm geom}(\lz)
 =\frac{1}{4G_4}\min_{\substack{\partial\gamma=\partial A_{\lz}\\
 \gamma\sim A_{\lz},\ z\geq\lz}}
 \operatorname{Area}(\gamma),
 \label{eq:RTl0}
\end{equation}
where $A_{\lz}$ is the copy of $A$ on the surface $z=\lz$, and the minimization runs over surfaces that stay in the region $z\geq\lz$. Taking the anchoring depth equal to the source-filter parameter is a choice. More generally one could anchor at $z_c=\kappa\lz$, and $\kappa=1$ is a calibration that we do not derive microscopically. In this sense no independent entanglement cutoff is introduced, but the identification itself is an assumption. Section~\ref{sec:holo} shows that the same surface $z=\lz$ also underlies the source filter, which gives the choice some motivation.

Equation~\eqref{eq:RTl0} is a prescription about a finite surface in a fixed geometry. It does not assert that source smearing has turned the parent CFT into a local theory with a finite cutoff. Finite-cutoff holography in the $T\bar T$ sense involves additional boundary dynamics and a modified dictionary \cite{MMV,Kraus}, which we do not specify. Throughout, $r$ and $\lz$ are compared as boundary coordinate lengths. Proper lengths on the anchoring surface are larger by the factor $L/\lz$, and $\lz$ is not a proper radial distance in AdS.

We first compute the entropy of a disk in $\mathrm{AdS}_4$. Let $A_{\lz}$ be the disk $0\leq r\leq R$ at $t=0$ on the surface $z=\lz$. By rotational symmetry the minimal surface can be described by a profile $z(r)$. Its induced metric is $(L^2/z^2)[(1+z'^2)\dd r^2+r^2\dd\theta^2]$, and the area functional is
\begin{equation}
 \operatorname{Area}[z]=2\pi L^2\int_0^R\dd r\,
 \frac{r}{z^2}\sqrt{1+z'(r)^2}.
 \label{eq:diskfunctional}
\end{equation}
It is to be minimized subject to $z(R)=\lz$ and $z'(0)=0$. The first condition anchors the surface on the edge of the disk; the second is smoothness on the symmetry axis. The resulting Euler--Lagrange equation is
\begin{equation}
 \frac{\dd}{\dd r}\left[
 \frac{rz'}{z^2\sqrt{1+z'^2}}\right]
 +\frac{2r}{z^3}\sqrt{1+z'^2}=0.
\end{equation}
Hemispheres $z^2+r^2=a^2$ centered on $z=0$ solve this equation, as for the ordinary RT disk. Conversely, every solution regular on the axis is such a hemisphere. Given the axis data $z(0)=z_*$ and $z'(0)=0$, the solution is unique, and the hemisphere of radius $z_*$ satisfies the same data. The anchoring condition $z(R)=\lz$ then fixes the radius:
\begin{equation}
 z(r)=\sqrt{a^2-r^2},\qquad a^2=R^2+\lz^2.
 \label{eq:diskcap}
\end{equation}
The surface is the part of the hemisphere of radius $a$ that lies above $z=\lz$. On it $z'=-r/z$, so $\sqrt{1+z'^2}=a/z$, and the area is an elementary integral:
\begin{equation}
 \begin{split}
 \operatorname{Area}(\gamma_A)
 &=2\pi L^2 a\int_0^R\frac{r\,\dd r}{(a^2-r^2)^{3/2}}\\
 &=2\pi L^2a\left[\frac{1}{\sqrt{a^2-r^2}}\right]_0^R\\
 &=2\pi L^2\left(\frac a{\lz}-1\right).
 \end{split}
\end{equation}
Dividing by $4G_4$ gives the exact geometric entropy:
\begin{equation}
 S_{\rm disk}^{\rm geom}(R;\lz)
 =\frac{\pi L^2}{2G_4}
 \left[\sqrt{1+\frac{R^2}{\lz^2}}-1\right].
 \label{eq:diskentropy}
\end{equation}
It is finite for every $R$ and every $\lz>0$, and it depends on the two lengths only through $R/\lz$.

For a well-resolved disk, $R\gg\lz$, expanding the square root gives
\begin{equation}
 S_{\rm disk}^{\rm geom}
 =\frac{L^2}{4G_4}\frac{2\pi R}{\lz}
 -\frac{\pi L^2}{2G_4}
 +\frac{\pi L^2\lz}{4G_4R}
 +O\!\left(\frac{L^2}{G_4}\frac{\lz^3}{R^3}\right).
 \label{eq:disklarge}
\end{equation}
The first term is proportional to the perimeter $2\pi R$ in units of $\lz$, which is the area law of a three-dimensional theory with resolution $\lz$. The constant term equals $-F$, where $F=\pi L^2/(2G_4)$ is the round-sphere free energy of the parent holographic CFT \cite{CHM,JKPS}. This constant is the universal, regulator-independent part of the disk entropy in three dimensions, and the finite resolution leaves it unchanged; it only adds corrections suppressed by powers of $\lz/R$.

\begin{figure*}[t]
 \centering
 \includegraphics[width=\textwidth]{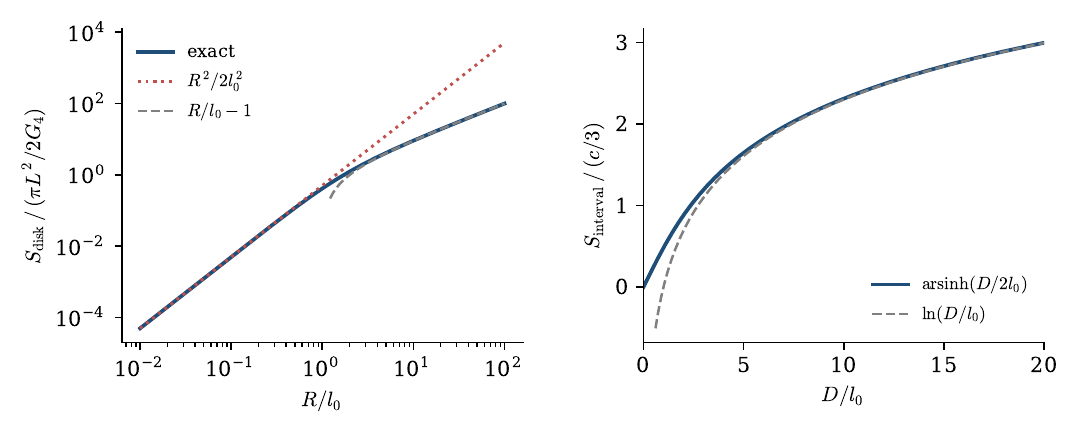}
 \caption{Left: exact disk entropy \eqref{eq:diskentropy} in units of $\pi L^2/(2G_4)$ (solid), with the subresolution behavior $R^2/2\lz^2$ (dotted) and the perimeter law with universal constant, $R/\lz-1$ (dashed). Right: $\mathrm{AdS}_3$ interval entropy \eqref{eq:intervalentropy} in units of $c/3$ (solid) compared with the logarithmic CFT form $\ln(D/\lz)$ (dashed).}
 \label{fig:entropy}
\end{figure*}

For a disk smaller than the resolution, $R\ll\lz$, the expansion runs the other way:
\begin{equation}
 S_{\rm disk}^{\rm geom}
 =\frac{\pi L^2R^2}{4G_4\lz^2}
 +O\!\left(\frac{L^2}{G_4}\frac{R^4}{\lz^4}\right).
\end{equation}
The entropy goes to zero smoothly, proportionally to the area $\pi R^2$ of the region (Fig.~\ref{fig:entropy}). Subresolution disks need not correspond to independently resolvable subsystems of a microscopic boundary theory, and this limit should be read as a property of the specified geometry.

It is instructive to compare the RT surface with the anchoring disk itself, which clarifies both the subresolution limit and the minimality of the hemisphere. The anchoring disk $A_{\lz}$ is itself one of the surfaces allowed in Eq.~\eqref{eq:RTl0}: it is homologous to itself and lies at $z\geq\lz$. Its induced metric at $t=0$, $z=\lz$ is $(L^2/\lz^2)(\dd r^2+r^2\dd\theta^2)$, so its area is $\operatorname{Area}(A_{\lz})=\pi L^2R^2/\lz^2$ and
\begin{equation}
 \frac{\operatorname{Area}(A_{\lz})}{4G_4}=\frac{\pi L^2R^2}{4G_4\lz^2}.
\end{equation}
This is exactly the leading term of the subresolution expansion. For $R\ll\lz$ the minimal surface is a shallow cap that barely rises above $z=\lz$ and hugs the anchoring disk, so its area approaches that of the disk and the entropy scales with the area of the boundary region rather than its perimeter. The hemisphere nevertheless has strictly smaller area for every $R>0$. Writing $y=\sqrt{1+R^2/\lz^2}>1$, the area of the RT surface is
\begin{equation}
 \frac{\operatorname{Area}(\gamma_A)}{2\pi L^2}=y-1,
\end{equation}
whereas that of the anchoring disk is
\begin{equation}
 \frac{\operatorname{Area}(A_{\lz})}{2\pi L^2}=\frac{R^2}{2\lz^2}=\frac{(y-1)(y+1)}{2}>y-1,
\end{equation}
where the inequality holds because $y+1>2$. The two areas coincide only in the limit $R\to0$.

Two regulator conventions should be distinguished. In the more common one, the RT surface is anchored on a disk of radius $R_\infty$ at $z=0$ and is merely cut off at $z=\lz$. Its area is that of the hemisphere of radius $R_\infty$ above $z=\lz$, which gives
\begin{equation}
 S_{\rm truncated}(R_{\infty};\lz)
 =\frac{\pi L^2}{2G_4}
 \left(\frac{R_{\infty}}{\lz}-1\right),
 \qquad R_{\infty}>\lz.
\end{equation}
The truncated surface meets $z=\lz$ on a circle of radius $\sqrt{R_\infty^2-\lz^2}$, not $R_\infty$. Equation~\eqref{eq:diskentropy} instead fixes the radius $R$ on the finite surface. The two prescriptions agree when $R_{\infty}=\sqrt{R^2+\lz^2}$, and they should not be compared at equal numerical radii.

Two further geometries confirm this picture. For a half-plane whose entangling line has coordinate length $H$, the minimal surface is the vertical plane that drops straight down from the line into the bulk. With the anchor at $z=\lz$, its area is
\begin{equation}
 \operatorname{Area}(\gamma_A)=L^2H\int_{\lz}^{\infty}\frac{\dd z}{z^2}
 =\frac{L^2H}{\lz},
\end{equation}
and the entropy per unit length of the entangling line is
\begin{equation}
 \frac{S_A^{\rm geom}}{H}=\frac{L^2}{4G_4\lz}.
 \label{eq:halfplane}
\end{equation}
This is ultraviolet finite and has the area-law form. An infinitely long entangling line still gives an infinite entropy, but this is an extensive infrared effect that $\lz$ is not meant to remove.

The same construction can be checked in $\mathrm{AdS}_3$, where the RT surface is a geodesic. Take a boundary interval of coordinate length $D$ with endpoints on the surface $z=\lz$. The minimal curve is an arc of the semicircle through the two endpoints, centered on $z=0$, and its length satisfies $\cosh(\mathcal L_\gamma/L)=1+D^2/(2\lz^2)$, or equivalently $\mathcal L_{\gamma}=2L\operatorname{arsinh}(D/2\lz)$. With the Brown--Henneaux central charge $c=3L/(2G_3)$ and $S=\mathcal L_\gamma/(4G_3)$,
\begin{equation}
 S_{\rm interval}^{\rm geom}
 =\frac c3\operatorname{arsinh}\!\left(\frac{D}{2\lz}\right)
 =\frac c3\ln\!\left[\frac{D+\sqrt{D^2+4\lz^2}}{2\lz}\right].
 \label{eq:intervalentropy}
\end{equation}
For $D\gg\lz$ this approaches the familiar CFT result $(c/3)\ln(D/\lz)$, with corrections of order $c\lz^2/D^2$, so $\lz$ plays the role of the usual short-distance cutoff in the logarithm. For $D\to0$ the entropy vanishes linearly, $S\simeq cD/(6\lz)$, instead of diverging to $-\infty$ as the CFT formula would. The same geodesic length appears in studies of cutoff $\mathrm{AdS}_3$ and $T\bar T$-deformed CFTs \cite{CCH}. Here it arises from the zero-point-length anchoring, and we attach no $T\bar T$ interpretation to it.

We close this section with two cautionary remarks. First, the entropy prescription and the source filter should not be conflated. The source map satisfies $W_{\lz}[0]=W_0[0]$: at zero source the theory is the parent CFT, and its vacuum density matrix is unchanged. Convolving operators across an entangling surface does not by itself define a new split of the degrees of freedom into those inside and outside a region, so a finite filtered two-point function does not prove that the von Neumann entropy is finite. Equation~\eqref{eq:RTl0} is an additional geometric prescription with its own anchoring data. We have not constructed a single boundary replica functional that yields both the finite-surface entropy and the Bessel-filtered correlator.

Second, at leading order the finite areas above realize $\lz$ as the only short-distance scale in the entropy calculation. They do not show that T-duality modifies the coefficient $1/(4G_4)$ or the RT functional itself. Higher-curvature corrections would require the corresponding entropy functional. At quantum order, bulk entanglement and gravitational renormalization also contribute, and smearing external sources does not regulate these terms.

\section{Mutual information and the finite-resolution connectivity transition}
\label{sec:mutualresolution}

The entropy of a single region does not determine how the finite-resolution
prescription resolves correlations between separated regions. A useful
extension is the mutual-information combination for two disjoint intervals.
Here we compute this quantity using precisely the geometric prescription
\eqref{eq:RTl0}, specialized to the vacuum of $\mathrm{AdS}_3$. The result
is conditional on the same finite-surface anchoring assumption as
Eq.~\eqref{eq:intervalentropy}; it is not derived from the source functional
$W_{\lz}[J]$. We work at leading classical order in the Einstein sector,
with $c=3L/(2G_3)$, and all interval lengths and separations below are
coordinate lengths on the common surface $z=\lz$. Besides the exact threshold for arbitrary interval lengths, we prove that a unique transition exists for
arbitrary intervals, establish an upper bound on the connectivity window
at fixed total interval length, and derive a subresolution scaling form
for unequal intervals. These results characterize the
finite-surface geometric prescription without assuming
a microscopic entropy interpretation for the filtered
boundary observables.

Let the endpoints at fixed time be
\begin{equation}
    \begin{gathered}
    x_1=0,\qquad x_2=D_1,\\
    x_3=D_1+h,\qquad x_4=D_1+h+D_2,
    \end{gathered}
\end{equation}
with $D_1,D_2,h>0$. The intervals are $A=[x_1,x_2]$ and $B=[x_3,x_4]$. Define
\begin{equation}
    s_{\lz}(w)\equiv\frac{c}{3}
    \operatorname{arsinh}\!\left(\frac{w}{2\lz}\right),
    \qquad w\geq0,
    \label{eq:MIinterval}
\end{equation}
which is the single-interval entropy \eqref{eq:intervalentropy} as a
function of the interval length $w$. On the constant-time hyperbolic
slice, there are two non-crossing geodesic pairings compatible with
homology. The disconnected pairing joins $(x_1,x_2)$ and $(x_3,x_4)$ and
gives
\begin{equation}
    S_{\rm disc}=s_{\lz}(D_1)+s_{\lz}(D_2),
\end{equation}
while the connected pairing joins $(x_1,x_4)$ and $(x_2,x_3)$ and gives
\begin{equation}
    S_{\rm conn}=s_{\lz}(h)+s_{\lz}(D_1+D_2+h).
    \label{eq:MIcandidates}
\end{equation}
Each geodesic segment remains in
$z\geq\lz$. A crossing pairing cannot be the global minimum: cutting
at the crossing and reconnecting gives a non-crossing competitor, which
can then be shortened to geodesics. Thus the classical RT rule
\cite{RT,RTlong} selects the smaller of the two candidates,
\begin{equation}
    S_{A\cup B}^{\rm geom}=\min\{S_{\rm disc},S_{\rm conn}\}.
\end{equation}
The geometric mutual information $I(A:B)=S_A+S_B-S_{A\cup B}$ is then
\begin{multline}
    I_{\lz}^{\rm geom}(A:B)=\max\big\{0,\,
        s_{\lz}(D_1)+s_{\lz}(D_2)\\
        -s_{\lz}(h)-s_{\lz}(D_1+D_2+h)\big\}.
    \label{eq:MIgeneral}
\end{multline}
This is finite even in the adjacent-interval limit $h\to0^+$ at fixed
$\lz>0$. A positive value selects the connected RT configuration.
The vanishing answer in the disconnected phase is a leading classical
statement; bulk quantum contributions can give non-zero mutual information.

We first show that a transition exists and is unique for arbitrary
intervals. To see this, define the difference between
the disconnected and connected candidates,
\begin{equation}
    \begin{split}
        F(D_1,D_2,h)={}&s_{\lz}(D_1)+s_{\lz}(D_2)\\
        &-s_{\lz}(h)-s_{\lz}(D_1+D_2+h).
    \end{split}
    \label{eq:MIbranchdifference}
\end{equation}
The key property is the shape of the interval function. Its first
derivative is positive,
\begin{equation}
    s_{\lz}'(w)=\frac{c}{3\sqrt{w^2+4\lz^2}}>0,
\end{equation}
so $s_{\lz}$ is increasing, while its second derivative is non-positive,
\begin{equation}
    s_{\lz}''(w)=
    -\frac{cw}{3(w^2+4\lz^2)^{3/2}}\leq0,
    \label{eq:MIconcavity}
\end{equation}
so $s_{\lz}$ is concave and strictly so for $w>0$. Since $s_{\lz}(0)=0$ and $s_{\lz}'$ is strictly
decreasing for positive argument, the branch difference is positive at
zero separation,
\begin{equation}
    F(D_1,D_2,0)
    =\int_0^{D_1}\dd u\,
    \left[s_{\lz}'(u)-s_{\lz}'(D_2+u)\right]>0.
\end{equation}
Moreover, $F$ decreases monotonically as the separation grows,
\begin{equation}
    \partial_h F
    =-s_{\lz}'(h)-s_{\lz}'(D_1+D_2+h)<0,
\end{equation}
and $F\to-\infty$ as $h\to\infty$. Hence every pair
$D_1,D_2>0$ has a unique critical separation
$h_c(D_1,D_2;\lz)$.

The threshold can also be obtained in closed form for arbitrary unequal
intervals. Write $\mathcal D=D_1+D_2$ and define
\begin{equation}
 \Theta=\frac12\left[
 \operatorname{arsinh}\!\left(\frac{D_1}{2\lz}\right)
 +\operatorname{arsinh}\!\left(\frac{D_2}{2\lz}\right)\right].
 \label{eq:MIunequalTheta}
\end{equation}
At the transition, equality of the connected and disconnected candidates
is equivalent to
\begin{equation}
 \begin{gathered}
 u+v=2\Theta,\\
 u=\operatorname{arsinh}\!\left(\frac{\mathcal D+h_c}{2\lz}\right),\qquad
 v=\operatorname{arsinh}\!\left(\frac{h_c}{2\lz}\right).
 \end{gathered}
\end{equation}
Writing $u=\Theta+b$ and $v=\Theta-b$, the difference of their
hyperbolic sines gives
\begin{equation}
 2\cosh\Theta\,\sinh b=\frac{\mathcal D}{2\lz},
 \qquad
 \sinh b=\frac{\mathcal D}{4\lz\cosh\Theta}.
\end{equation}
Define the length
\begin{equation}
 \mathcal Q=\sqrt{\mathcal D^2+16\lz^2\cosh^2\Theta}.
\end{equation}
Substitution into $h_c=2\lz\sinh(\Theta-b)$ then yields the exact result
\begin{equation}
 \displaystyle
 h_c(D_1,D_2;\lz)=\frac{\mathcal Q\tanh\Theta-\mathcal D}{2}.
 \label{eq:MIcriticalunequalExact}
\end{equation}
The positive root is selected because $\cosh b>0$, and the previously
proved uniqueness identifies this expression with the physical crossing
of the two geometric candidates. The connected configuration is preferred
for $0\leq h<h_c$, and the disconnected one for $h>h_c$.
For $D_1=D_2=D$, one has $\Theta=\operatorname{arsinh}(D/2\lz)$,
and Eq.~\eqref{eq:MIcriticalunequalExact} reduces to
Eq.~\eqref{eq:MIcritical}. At fixed positive $D_1,D_2$, its
conformal-boundary limit is
\begin{equation}
 h_c(D_1,D_2;0)=
 \frac{\sqrt{\mathcal D^2+4D_1D_2}-\mathcal D}{2}.
 \label{eq:MIcriticalunequalIR}
\end{equation}
For $\mathcal D\ll\lz$ at fixed positive ratio $D_1/D_2$, it gives
$h_c=D_1D_2\mathcal D/(16\lz^2)
[1+O(\mathcal D^2/\lz^2)]$, in agreement with the independent
expansion below. Thus numerical root finding is not required to locate
the transition for unequal intervals either. All these thresholds refer
to the same finite-surface geometric prescription.

A useful comparison follows by fixing the total
interval length $\mathcal D=D_1+D_2$. Concavity gives
\begin{equation}
    s_{\lz}(D_1)+s_{\lz}(D_2)
    \leq 2s_{\lz}(\mathcal D/2),
    \label{eq:MIequaloptimal}
\end{equation}
where equality holds only for $D_1=D_2$.
The connected candidate,
$s_{\lz}(h)+s_{\lz}(\mathcal D+h)$, is independent
of the partition of $\mathcal D$. Consequently, the branch difference
for any partition is bounded by its equal-interval value, and the mutual
information obeys
\begin{equation}
    I_{\lz}^{\rm geom}(D_1,D_2,h)
    \leq
    I_{\lz}^{\rm geom}(\mathcal D/2,\mathcal D/2,h).
    \label{eq:MIinfobound}
\end{equation}
Because $F$ decreases with $h$, the zero of the unequal branch difference
cannot lie beyond that of the equal one, so the critical separations obey
\begin{equation}
    h_c(D_1,D_2;\lz)
    \leq
    h_c(\mathcal D/2,\mathcal D/2;\lz).
    \label{eq:MIasymmetrybound}
\end{equation}
Thus equal intervals maximize both the geometric
mutual information and the connectivity window at
fixed total interval length. The equal-interval
transition derived below therefore provides an
upper bound for all partitions of the same
$\mathcal D$.

\begin{figure*}[t]
    \centering
    \includegraphics[width=\textwidth]{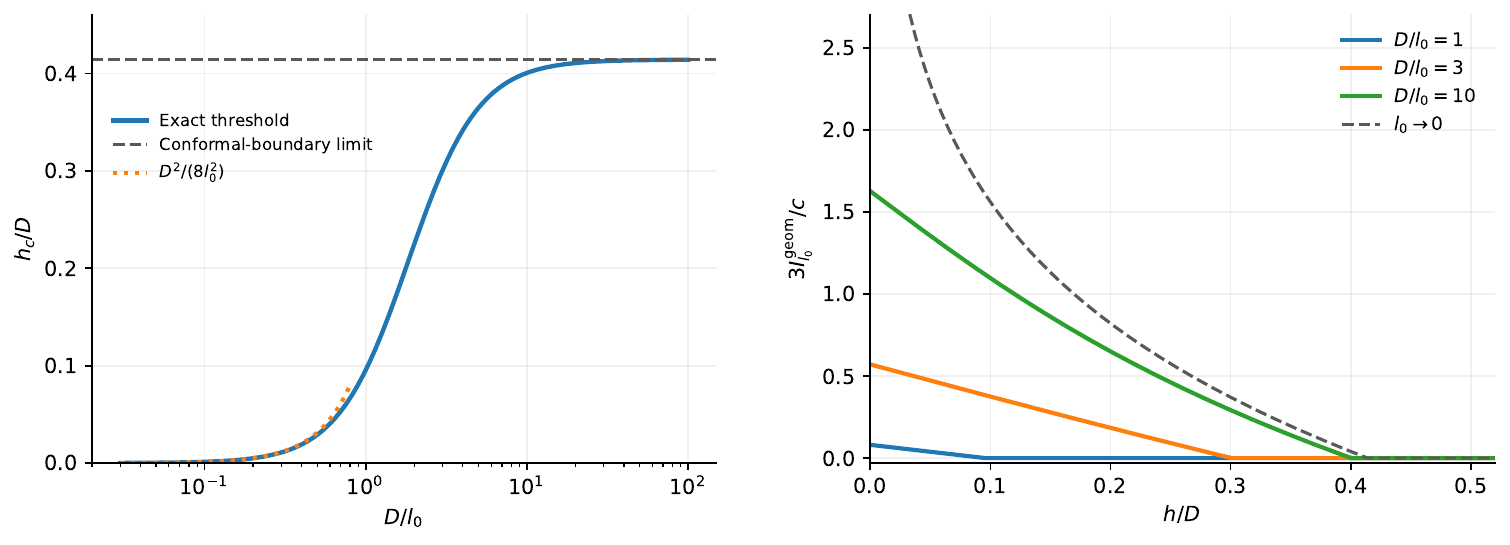}
    \caption{Mutual information under the finite-surface geometric
        prescription. Left: the exact equal-interval transition
        $h_c/D$ as a function of $D/\lz$, compared with the conformal-boundary
        limit $\sqrt{2}-1$ and the subresolution asymptote
        $D^2/(8\lz^2)$. Right: $3I_{\lz}^{\rm geom}/c$ as a function of the
        gap ratio $h/D$ for several resolutions, compared with the
        $\lz\to0$ result. Finite resolution shifts the connectivity threshold
        to smaller separations. These are leading classical geometric
        quantities, not entropies derived from the Bessel-filtered source
        functional.}
    \label{fig:MIresolution}
\end{figure*}

For equal intervals the general threshold simplifies to an elementary expression, which we also derive directly. For $D_1=D_2=D$, introduce $x=D/(2\lz)$ and
$y=h/(2\lz)$. The connected branch, in units of $c/3$, is controlled by
the function
\begin{equation}
    f(x,y)=2\operatorname{arsinh}x
    -\operatorname{arsinh}y-\operatorname{arsinh}(2x+y),
\end{equation}
in terms of which the mutual information reads
\begin{equation}
    I_{\lz}^{\rm geom}=\frac{c}{3}\max\{0,f(x,y)\}.
    \label{eq:MIequal}
\end{equation}
For every $x>0$ one has $f(x,0)>0$, while $f$ decreases monotonically
with the separation,
\begin{equation}
    \partial_y f=-\frac{1}{\sqrt{1+y^2}}
    -\frac{1}{\sqrt{1+(2x+y)^2}}<0,
\end{equation}
and tends to $-\infty$ as $y\to\infty$. There is therefore a unique connectivity transition. It can be found
without numerically minimizing any surface. Write $a=\operatorname{arsinh}x$
and, at the transition,
\begin{equation}
    \operatorname{arsinh}(2x+y)=a+b,\qquad
    \operatorname{arsinh}y=a-b.
\end{equation}
Taking the difference and the sum of the hyperbolic sines of these two
relations gives
\begin{equation}
    \sinh b=\frac{x}{\sqrt{1+x^2}},\qquad
    x+y=x\cosh b.
\end{equation}
Eliminating $b$ between these two relations and returning to the
variables $D=2\lz x$ and $h=2\lz y$, we obtain the critical separation
in closed form,
\begin{equation}
    h_c(D,\lz)=D\left[
        \sqrt{\frac{4\lz^2+2D^2}{4\lz^2+D^2}}-1\right].
    \label{eq:MIcritical}
\end{equation}
The connected configuration dominates for $0\leq h<h_c$, and the
disconnected one dominates for $h>h_c$. In particular,
\begin{equation}
    0<\frac{h_c}{D}<\sqrt{2}-1
    \qquad (D,\lz>0).
    \label{eq:MIbounds}
\end{equation}
At fixed $D$, increasing $\lz$ decreases $h_c$: writing
$q=4\lz^2/D^2$, the derivative of the square root in
Eq.~\eqref{eq:MIcritical} with respect to $q$ is
$-\{2(q+1)^2\sqrt{(q+2)/(q+1)}\}^{-1}<0$.
Thus finite anchoring reduces the coordinate separation over which the
classical connected configuration is favored. The threshold depends on
$D/\lz$, not only on the ratio $h/D$, reflecting the additional resolution
scale (Fig.~\ref{fig:MIresolution}).

The exact results simplify in two limits. For $D,h\gg\lz$, expanding
the difference of the two candidate entropies gives
\begin{align}
    \frac{3}{c}(S_{\rm disc}-S_{\rm conn})
    ={}&\ln\frac{D^2}{h(2D+h)}
    \nonumber\\
    &+\lz^2\left[\frac{2}{D^2}-\frac{1}{h^2}
    -\frac{1}{(2D+h)^2}\right]
    \nonumber\\
    &+O\!\left(\lz^4\left[D^{-4}+h^{-4}+(2D+h)^{-4}\right]\right).
    \label{eq:MIlarge}
\end{align}
In the limit $\lz\to0$ at fixed non-zero $D,h$, the usual classical
vacuum result is recovered,
\begin{equation}
    I_0^{\rm geom}=\frac{c}{3}\max\!\left\{0,
    \ln\frac{D^2}{h(2D+h)}\right\},
\end{equation}
which vanishes beyond the familiar conformal-boundary threshold
\begin{equation}
    h_c(D,0)=(\sqrt{2}-1)D.
\end{equation}
The leading shift of the transition follows directly from
Eq.~\eqref{eq:MIcritical}:
\begin{equation}
    h_c=(\sqrt{2}-1)D-\frac{\sqrt{2}\,\lz^2}{D}
    +O\!\left(\frac{\lz^4}{D^3}\right),
    \qquad D\gg\lz.
    \label{eq:MIcriticalIR}
\end{equation}
The correction is negative, so finite resolution narrows the window
in which the connected configuration dominates.

In the opposite, subresolution limit, expanding
Eq.~\eqref{eq:MIcritical} in powers of $D^2/\lz^2$ gives
\begin{equation}
    h_c=\frac{D^3}{8\lz^2}
        -\frac{5D^5}{128\lz^4}
        +O\!\left(\frac{D^7}{\lz^6}\right),
        \qquad D\ll\lz.
    \label{eq:MIcriticalUV}
\end{equation}
The connected window is then much narrower than either interval.
Since the branch difference decreases with $h$, the supremum over
disjoint configurations is approached as $h\to0^+$. Denoting this continuous adjacent-interval
limit by $I_{\lz}^{\rm geom}(D,0)$, we obtain
\begin{align}
    I_{\lz}^{\rm geom}(D,0)
    &=\frac{c}{3}\left[
    2\operatorname{arsinh}\!\left(\frac{D}{2\lz}\right)
    -\operatorname{arsinh}\!\left(\frac{D}{\lz}\right)\right]
    \nonumber\\
    &=\frac{cD^3}{24\lz^3}
    +O\!\left(\frac{cD^5}{\lz^5}\right).
    \label{eq:MIadjacent}
\end{align}
The cancellation of the terms linear in $D/\lz$ explains both the cubic
suppression of adjacent-interval mutual information and the small
critical gap in Eq.~\eqref{eq:MIcriticalUV}. At any fixed positive ratio
$h/D$, sufficiently small $D/\lz$ therefore selects the disconnected
classical configuration.

The narrow connected window has a simple scaling
form that extends to unequal intervals. Let
$\mathcal D=D_1+D_2$, take $\mathcal D/\lz\ll1$
at fixed positive ratio $D_1/D_2$, and consider
gaps $h=O(\mathcal D^3/\lz^2)$. Expanding
Eq.~\eqref{eq:MIbranchdifference} gives
\begin{equation}
    F(D_1,D_2,h)
    =
    \frac{cD_1D_2\mathcal D}{48\lz^3}
    -\frac{ch}{3\lz}
    +O\!\left(\frac{c\mathcal D^5}{\lz^5}\right).
    \label{eq:MIsmallunequal}
\end{equation}
The linear terms in the interval lengths cancel,
while the gap contributes already at linear order.
The critical separation therefore satisfies
\begin{equation}
    h_c(D_1,D_2;\lz)
    =
    \frac{D_1D_2\mathcal D}{16\lz^2}
    \left[1+O\!\left(\frac{\mathcal D^2}{\lz^2}\right)\right].
    \label{eq:MIcriticalunequalUV}
\end{equation}

Define the leading critical scale
$h_* = D_1D_2\mathcal D/(16\lz^2)$ and the
dimensionless gap $\eta=h/h_*$. For fixed $\eta$,
\begin{equation}
    \frac{48\lz^3}{cD_1D_2\mathcal D}
    I_{\lz}^{\rm geom}
    =
    \max\{0,1-\eta\}
    +O\!\left(\frac{\mathcal D^2}{\lz^2}\right).
    \label{eq:MIscaling}
\end{equation}
Thus the rescaled mutual information approaches a
common piecewise-linear profile in this limit.
For $D_1=D_2=D$, this reproduces both the leading
critical gap in Eq.~\eqref{eq:MIcriticalUV} and
the adjacent-interval amplitude in
Eq.~\eqref{eq:MIadjacent}.

The effect of asymmetry is particularly transparent
if we parametrize the two interval lengths as
\begin{equation}
    D_1=\frac{\mathcal D}{2}(1+\delta),
    \qquad
    D_2=\frac{\mathcal D}{2}(1-\delta),
    \qquad |\delta|<1.
\end{equation}
At fixed $\delta$ one has $D_1D_2=\mathcal D^2(1-\delta^2)/4$, and
Eq.~\eqref{eq:MIcriticalunequalUV} becomes
\begin{equation}
    h_c
    =
    \frac{\mathcal D^3}{64\lz^2}(1-\delta^2)
    \left[1+O\!\left(\frac{\mathcal D^2}{\lz^2}\right)\right].
    \label{eq:MIasymmetryUV}
\end{equation}
The factor $1-\delta^2$ quantifies the reduction
of the subresolution connectivity window by
interval-size asymmetry.

\begin{figure*}[t]
    \centering
    \includegraphics[width=\textwidth]{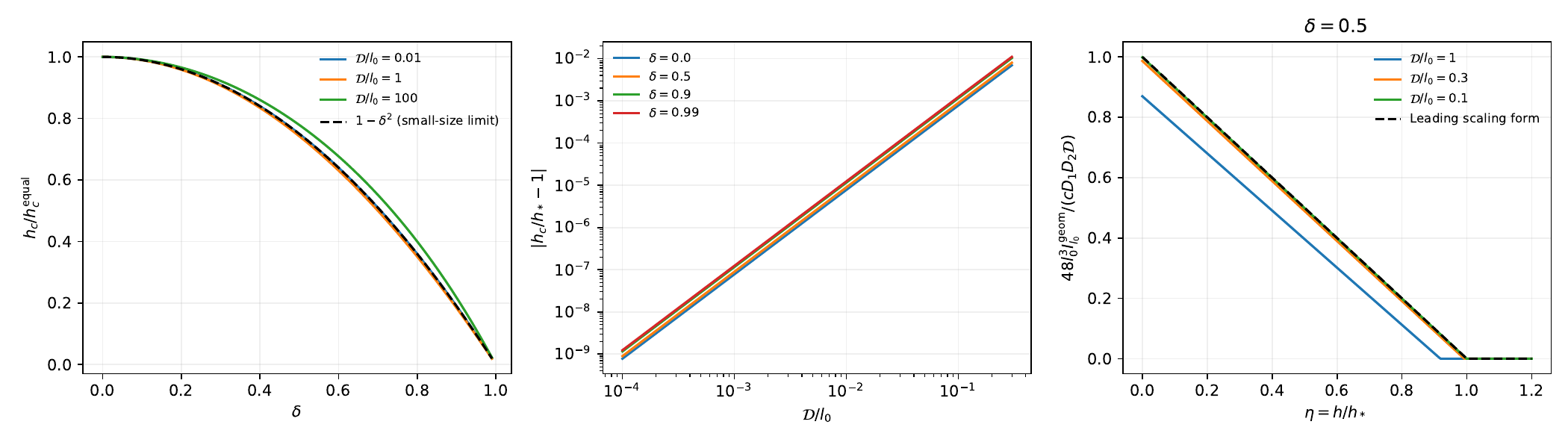}
    \caption{
        Unequal-interval connectivity and subresolution scaling
        under the finite-surface geometric prescription.
        We write $\mathcal D=D_1+D_2$,
        $\delta=(D_1-D_2)/\mathcal D$, and
        $h_*=D_1D_2\mathcal D/(16\lz^2)$.
        Left: the critical separation normalized by
        $h_c^{\rm equal}\equiv
        h_c(\mathcal D/2,\mathcal D/2;\lz)$, as a function
        of asymmetry for $\mathcal D/\lz=0.01,1,100$.
        The dashed curve is the subresolution limit
        $h_c/h_c^{\rm equal}\to1-\delta^2$.
        Only $\delta\geq0$ is shown because exchanging
        the intervals sends $\delta\to-\delta$ without
        changing the result.
        Middle: the relative error $|h_c/h_*-1|$ for
        $\delta=0,0.5,0.9,0.99$, exhibiting the quadratic
        scaling predicted by
        Eq.~\eqref{eq:MIcriticalunequalUV}.
        Right: the rescaled geometric mutual information
        $48\lz^3 I_{\lz}^{\rm geom}/(cD_1D_2\mathcal D)$
        versus $\eta=h/h_*$ at $\delta=0.5$ for
        $\mathcal D/\lz=1,0.3,0.1$.
        The dashed curve is the limiting profile
        $\max\{0,1-\eta\}$ in Eq.~\eqref{eq:MIscaling}.
        All quantities are evaluated at leading classical order.
    }
    \label{fig:MIunequal}
\end{figure*}

We have also checked the unequal-interval results numerically, by directly
evaluating Eq.~\eqref{eq:MIgeneral} and numerically
solving $F(D_1,D_2,h_c)=0$. Across 549 combinations
of total interval length and asymmetry, with
$10^{-4}\leq\mathcal D/\lz\leq10^3$ and
$|\delta|\leq0.99$, no violation of the
critical-separation bound in
Eq.~\eqref{eq:MIasymmetrybound} was found.
The corresponding mutual-information bound
\eqref{eq:MIinfobound} was also satisfied in 3843 sampled configurations.
For fixed asymmetries
$\delta=0,0.5,0.9,0.99$, high-precision
calculations confirmed that $h_c/h_*\to1$
as $\mathcal D/\lz\to0$.
Both the relative error in the leading
critical-gap approximation and the sampled
absolute error in the rescaled profile
\eqref{eq:MIscaling} exhibited the expected
$O((\mathcal D/\lz)^2)$ scaling.
The branch difference was evaluated with
60-decimal-digit arithmetic to control
cancellation at small interval sizes.
These checks support the analytic results over the sampled
parameter ranges. Figure~\ref{fig:MIunequal} illustrates the
reduction of the connectivity window by interval-size
asymmetry, the quadratic convergence of the leading
critical-gap approximation, and the approach to the
rescaled profile in Eq.~\eqref{eq:MIscaling}.

We emphasize the scope of this result. The calculation extends the single-interval geometry of
Eq.~\eqref{eq:intervalentropy}, whose finite-cutoff counterpart is already
known \cite{CCH}, to competing RT configurations. Neither the geodesic
construction nor the use of mutual information is specific to
T-duality. The dependence on $\lz$ here follows from the calibration
$z_c=\lz$; choosing $z_c=\kappa\lz$ replaces $\lz$ by $\kappa\lz$
throughout this section. In particular, $h_c$ is a connectivity threshold
of this classical geometric prescription, not a correlation length of
the source-filtered theory. A disconnected RT configuration does not
imply that all boundary correlators vanish, and no microscopic
factorization of the smeared observables has been assumed.

\section{Extensions and limits}
\label{sec:extensions}

Nothing in the construction is specific to the massless scalar. For a scalar of mass $m$ in Euclidean $\mathrm{AdS}_{d+1}$ with standard quantization, the dual operator has dimension
\begin{equation}
 \Delta=\frac d2+\nu,\qquad
 \nu=\sqrt{\frac{d^2}{4}+m^2L^2},
\end{equation}
assuming $m^2$ respects the Breitenlohner--Freedman bound and appropriate boundary conditions are imposed. The source is now the coefficient of $z^{d-\Delta}$ in the near-boundary expansion. Replacing it by $\PP_{\lz}J$ changes neither the field equation nor its indicial roots, and the argument of Sec.~\ref{sec:sources} goes through unchanged:
\begin{equation}
 \mathcal G_{\lz}(p)=\PP_{\lz}(p)^2\,\mathcal G_0(p).
\end{equation}
This holds for whatever $\mathcal G_0$ the parent scheme produces, including logarithms and contact terms.

The same reasoning extends to interacting theories. If the parent theory includes bulk interactions, the generating functional is still $W_0[\PP_{\lz}J]$, and differentiating $n$ times with respect to $J$ gives, by the chain rule,
\begin{equation}
 \begin{split}
 &\langle\OO_{\lz}(p_1)\cdots\OO_{\lz}(p_n)\rangle_c\\
 &\qquad=\left[\prod_{a=1}^n\PP_{\lz}(p_a)\right]
 \langle\OO(p_1)\cdots\OO(p_n)\rangle_c.
 \end{split}
 \label{eq:npoint}
\end{equation}
This identity does not rely on the semiclassical approximation. In Witten-diagram language it places one filter on each external leg and none on internal propagators. It therefore does not make bulk loops ultraviolet finite; that question is taken up in Sec.~\ref{sec:bulkloops}.

We next distinguish the construction from two related ideas, a radial cutoff and a regular black hole. First, filtering the source by $\PP_{\lz}$ is different from moving the boundary to a finite depth $z_c$. The ordinary solution sampled at $z_c$ has transfer function $(1+pz_c)e^{-pz_c}$. Its small-momentum expansion is $1-p^2z_c^2/2+p^3z_c^3/3+\dots$. It has no $p^2\ln p$ term, although the $p^3$ term is nonanalytic in $p^2$. The Bessel filter has a different low-momentum structure and also differs at large momentum. Beyond this, a finite-cutoff generating functional is defined by a different boundary-value problem altogether. The identification $z_c=\lz$ used in Sec.~\ref{sec:entropy} is therefore an additional assignment and not a consequence of the source map. Section~\ref{sec:holo} sharpens this statement. Although $\PP_{\lz}$ is not the transfer function of the probe at $z_c$, it is exactly the transfer function of a different field, with $\nu=1$, evaluated at $z=\lz$. The surface $z=\lz$ thus enters both constructions, while the entropy prescription remains an independent assumption.

Second, the metric remains Eq.~\eqref{eq:metric}, and no zero-point-length black hole has been produced. A genuine distributed source in the bulk would require an inhomogeneous field equation such as
\begin{equation}
 (\nabla^2-m^2)\phi=\mathcal I_{\lz}(X),
\end{equation}
with an AdS-covariant source profile $\mathcal I_{\lz}$. Its solution is the sum of a homogeneous piece and the source convolved with the Dirichlet Green function $G_D$,
\begin{equation}
 \phi=\phi_{\rm hom}+
 \int\dd^{d+1}X'\sqrt{g'}\,G_D(X,X')\,\mathcal I_{\lz}(X').
\end{equation}
The flat three-dimensional profile \eqref{eq:rho3} cannot simply be inserted here, because its radial embedding and its normalization with respect to proper volume would have to be specified. A covariant candidate is constructed in Sec.~\ref{sec:bulkloops}. A backreacted geometry would further require a consistent matter action whose conserved stress tensor satisfies the Einstein constraints.

Third, $\lz$ is a boundary coordinate resolution and not automatically a bulk proper length. Under the AdS dilatation $(z,x)\mapsto(\lambda z,\lambda x)$ it must scale as $\lz\mapsto\lambda\lz$ to map the family of probes into itself. Holding $\lz$ fixed breaks the scale covariance of the measured correlators, as it should for a physical resolution. Identifying $\lz$ with a compactification radius or with $\sqrt{\alpha'}$ \cite{SSP,FSP} would require an explicit string embedding and a normalization dictionary.

\section{Holographic origin of the filter}
\label{sec:holo}
So far the Bessel filter has been imposed from outside, as an input borrowed from zero-point-length models. In this section we show that it has a direct holographic meaning: it is the bulk-to-boundary propagator of a particular AdS field, evaluated at radial depth $\lz$.

Consider a scalar of mass $m$ in Euclidean $\mathrm{AdS}_{d+1}$, with the metric \eqref{eq:metric} extended to $d$ boundary dimensions. We derive the identity in seven steps.

\emph{Step 1: radial equation.} After a Fourier transform along the boundary, the field equation becomes
\begin{equation}
 \phi''-\frac{d-1}{z}\phi'-\left(p^2+\frac{m^2L^2}{z^2}\right)\phi=0 ,
\end{equation}
which reduces to Eq.~\eqref{eq:eom} for $d=3$ and $m=0$. The substitution $\phi=z^{d/2}f(pz)$ turns it into the modified Bessel equation $x^2f''+xf'-(x^2+\nu^2)f=0$ with $\nu=\sqrt{d^2/4+m^2L^2}$. Of its two solutions, $I_\nu$ grows exponentially at large $x$, so the solution regular in the interior is $f=K_\nu$.

\emph{Step 2: near-boundary behavior.} For $\nu>0$, $K_\nu(x)\simeq\tfrac12\Gamma(\nu)(2/x)^\nu$ as $x\to0$. Hence
\begin{equation}
 z^{d/2}K_\nu(pz)\simeq2^{\nu-1}\Gamma(\nu)\,p^{-\nu}z^{d-\Delta},\qquad \Delta=\frac d2+\nu ,
\end{equation}
which is the source falloff $z^{d-\Delta}$ with a momentum-dependent coefficient.

\emph{Step 3: reduced transfer function.} Dividing by this coefficient normalizes the solution to the source, $\phi(z,p)=z^{d-\Delta}\chi_\nu(z,p)\,j(p)$, with
\begin{equation}
 \chi_\nu(z,p)=\frac{(pz)^\nu K_\nu(pz)}{2^{\nu-1}\Gamma(\nu)},\qquad \chi_\nu\to1\ \ (z\to0).
 \label{eq:chinu}
\end{equation}
The function $\chi_\nu$ tells how a boundary mode of momentum $p$ propagates to depth $z$, once the trivial power $z^{d-\Delta}$ is stripped off. It depends on $\nu$ but not on $d$.

\emph{Step 4: check against the probe.} For $\nu=3/2$ the Bessel function is elementary, $K_{3/2}(x)=\sqrt{\pi/2x}\,e^{-x}(1+1/x)$, and $2^{1/2}\Gamma(3/2)=\sqrt{\pi/2}$. Equation~\eqref{eq:chinu} then gives $\chi_{3/2}=(1+pz)e^{-pz}$, which is the transfer function \eqref{eq:bulkF} of the massless probe.

\emph{Step 5: the identity.} For $\nu=1$ the normalization factor is $2^0\Gamma(1)=1$, and Eq.~\eqref{eq:chinu} at $z=\lz$ is exactly the zero-point-length filter:
\begin{equation}
 \PP_{\lz}(p)=\lz p\,K_1(\lz p)=\chi_1(\lz,p).
 \label{eq:chi1}
\end{equation}

\emph{Step 6: Schwinger representation.} The identity extends to the proper-time representations of Sec.~\ref{sec:kernel}. The integral $\int_0^\infty\dd s\,s^{\mu-1}e^{-as-b/s}=2(b/a)^{\mu/2}K_\mu(2\sqrt{ab})$ with $\mu=-\nu$, $a=p^2$ and $b=z^2/4$ gives $\int_0^\infty\dd s\,s^{-\nu-1}e^{-sp^2-z^2/4s}=2^{1+\nu}p^\nu z^{-\nu}K_\nu(pz)$, and therefore
\begin{equation}
 \chi_\nu(z,p)=\frac{z^{2\nu}}{4^\nu\Gamma(\nu)}\int_0^\infty\dd s\,s^{-\nu-1}e^{-sp^2-z^2/4s}.
 \label{eq:chiSchwinger}
\end{equation}
At $p=0$ the integral equals $\Gamma(\nu)(4/z^2)^\nu$, which confirms $\chi_\nu(z,0)=1$. For $\nu=1$ and $z=\lz$, Eq.~\eqref{eq:chiSchwinger} coincides with the proper-time representation \eqref{eq:Pproper} of $\PP_{\lz}$. The zero-point-length weight $e^{-\lz^2/4s}$ is thus the Gaussian factor $e^{-z^2/4s}$ that describes radial propagation to depth $z=\lz$.

\emph{Step 7: position space.} We Fourier transform Eq.~\eqref{eq:chiSchwinger} using the Gaussian integral
\begin{equation*}
 \int\frac{\dd^dp}{(2\pi)^d}\,e^{ip\cdot x}e^{-sp^2}=(4\pi s)^{-d/2}e^{-x^2/4s}.
\end{equation*}
Substituting $u=1/s$ gives a Gamma-function integral:
\begin{equation}
 \begin{split}
 \chi_\nu(z,x)&=\frac{z^{2\nu}}{4^\nu\Gamma(\nu)(4\pi)^{d/2}}\int_0^\infty\dd u\,u^{\Delta-1}e^{-u(x^2+z^2)/4}\\
 &=\frac{\Gamma(\Delta)}{\pi^{d/2}\Gamma(\nu)}\frac{z^{2\Delta-d}}{(x^2+z^2)^\Delta}
 =z^{\Delta-d}K_\Delta(z,x).
 \end{split}
\end{equation}
In the last step we have introduced the standard bulk-to-boundary propagator
\begin{equation}
 K_\Delta(z,x)=\frac{\Gamma(\Delta)}{\pi^{d/2}\Gamma(\Delta-\frac d2)}\,\frac{z^\Delta}{(z^2+x^2)^\Delta}
\end{equation}
which is normalized so that $z^{\Delta-d}K_\Delta\to\delta^{(d)}(x)$ as $z\to0$. Setting $\nu=1$, $d=n$ (so that $\Delta=n/2+1$) and $z=\lz$ reproduces the position-space kernel of Sec.~\ref{sec:kernel}:
\begin{equation}
 \begin{split}
 \rho^{(n)}_{\lz}(x)&=\lz^{\Delta-n}K_\Delta(\lz,x)\Big|_{\Delta=n/2+1}\\
 &=\frac{\Gamma\!\left(\frac{n+2}{2}\right)}{\pi^{n/2}}\frac{\lz^2}{(x^2+\lz^2)^{(n+2)/2}}.
 \end{split}
\end{equation}
The properties of $\rho^{(n)}_{\lz}$ found earlier, namely positivity, unit normalization, the $\delta$-function limit and the power-law tail, are those of $K_\Delta$.

Three remarks are in order. (i) The value $\nu=1$ is fixed by the zero-point-length prescription and not by the bulk dimension. Padmanabhan's kernel $\int_0^\infty\dd s\,e^{-sp^2-\lz^2/4s}=(\lz/p)K_1(\lz p)$ has the same form in any number of dimensions, and it always selects the order-one Bessel function. (ii) For the $d=3$ boundary of this paper the corresponding channel is a scalar with
\begin{equation}
 m_\Phi^2L^2=\nu^2-\tfrac94=-\tfrac54,\qquad \Delta_\Phi=\tfrac52 ,
\end{equation}
which is tachyonic but lies above the Breitenlohner--Freedman bound $m^2L^2=-\tfrac94$, and is therefore stable in AdS. (iii) The result can be summarized as follows: \emph{a boundary probe with zero-point-length resolution $\lz$ is holographically a probe sourced from bulk depth $z=\lz$ through a $\nu=1$ channel.} This accounts for several features derived separately above. The positivity, normalization and heavy tails of $\rho_{\lz}$ are those of a bulk-to-boundary propagator. The time non-locality behind Eq.~\eqref{eq:RPfail} reflects the fact that Euclidean propagation from $z=\lz$ to the boundary is not local in boundary time. The factor $\PP_{\lz}^2$ in Eq.~\eqref{eq:mainG} corresponds to one bulk-to-boundary leg per insertion, consistent with Eq.~\eqref{eq:npoint}.

\section{A two-field auxiliary transfer prescription}
\label{sec:twofield}
The identity \eqref{eq:chi1} represents the filter by radial propagation of an auxiliary field. We use this as a prescribed transfer construction: the action below supplies two free propagation problems, while an additional identification specifies how their data are related. The action alone does not derive that identification. Consider, in Euclidean $\mathrm{AdS}_4$,
\begin{equation}
 \begin{split}
 I={}&\frac{\mathcal N}{2}\int\dd^4X\sqrt g\,(\partial\phi)^2\\
 &+\frac{\mathcal N_\Phi}{2}\int\dd^4X\sqrt g\left[(\partial\Phi)^2+m_\Phi^2\Phi^2\right],
 \end{split}
\end{equation}
with $m_\Phi^2L^2=-\tfrac54$, where $\phi$ is the massless probe and $\Phi$ is an auxiliary $\nu=1$ field. The experimental source is carried by the auxiliary field, $\Phi(z,x)\to z^{1/2}J(x)$ as $z\to0$, where $z^{1/2}=z^{d-\Delta_\Phi}$ is its source falloff.

\emph{Auxiliary solution.} By Eq.~\eqref{eq:chinu} with $d=3$, $\nu=1$ and $\Delta_\Phi=5/2$, the regular solution is
\begin{equation}
 \Phi(z,p)=z^{1/2}\chi_1(z,p)J(p)=z^{1/2}\,pz\,K_1(pz)\,J(p).
\end{equation}

\emph{Identification.} We prescribe the Dirichlet datum of the probe to be the boundary-normalized value of $\Phi$ on the surface $z=\lz$:
\begin{equation}
 j(x)\equiv\lz^{-(d-\Delta_\Phi)}\Phi(\lz,x)=\lz^{-1/2}\Phi(\lz,x),
\end{equation}
hence
\begin{equation}
 j(p)=\lz pK_1(\lz p)J(p)=\PP_{\lz}(p)J(p).
\end{equation}
This is the source map \eqref{eq:sourcemap}. In position space it reads $j=\rho^{(3)}_{\lz}*J$ with $\rho^{(3)}_{\lz}=\lz^{-1/2}K_{5/2}(\lz,\cdot)$: the source $J$ propagates from the boundary to depth $\lz$ through the auxiliary field, and the value it reaches there becomes the boundary condition of the probe.

\emph{Probe and correlator.} With this boundary condition the probe solution is Eq.~\eqref{eq:mainbulk}, unchanged. If the auxiliary solution is first fixed as external transfer data and its sampled value is then assigned as the probe source, the composed on-shell functional is
\begin{equation}
 W[J]=W_\Phi[J]+W_0[\PP_{\lz}J],
 \label{eq:Wsplit}
\end{equation}
and the second term reproduces Eq.~\eqref{eq:mainG}. In position space the filtered correlator is the ordinary one with a bulk-to-boundary leg attached to each insertion:
\begin{equation}
 \begin{split}
 \langle\OO_{\lz}(x_1)\OO_{\lz}(x_2)\rangle
 ={}&\frac1{\lz}\int\dd^3y_1\,\dd^3y_2\,K_{5/2}(\lz,x_1-y_1)\\
 &\times K_{5/2}(\lz,x_2-y_2)\,\mathcal G_0(y_1-y_2).
 \end{split}
\end{equation}

It remains to discuss the first term in Eq.~\eqref{eq:Wsplit}. The term $W_\Phi$ is the two-point function of the $\Delta_\Phi=5/2$ operator $\OO_\Phi$, which $J$ sources as well. We compute it in three steps.

\emph{Step 1: expansion.} Inserting $xK_1(x)=1+\tfrac{x^2}{2}\left[\ln\tfrac x2+\gamma_E-\tfrac12\right]+O(x^4\ln x)$ into the auxiliary solution gives
\begin{equation}
 \Phi(z,p)=z^{1/2}J+z^{5/2}\,\frac{p^2}{2}\left[\ln p+\ln\frac z2+\gamma_E-\frac12\right]J+\dots
\end{equation}
Because $\nu=1$ is an integer, the two falloffs $z^{1/2}$ and $z^{5/2}$ differ by an integer power of $z^2$ and a logarithm appears. The $z^{5/2}\ln z$ term is fixed locally by the source, and so are the constants multiplying $p^2$. The non-local part of the normalizable coefficient is $\tfrac{p^2}{2}\ln p\,J$.

\emph{Step 2: response.} We use the standard relation $\langle\OO\rangle=(2\Delta-d)\,\mathcal N L^2\,\phi_{(2\nu)}$ up to local terms, where $\phi_{(2\nu)}$ is the normalizable coefficient. For $\nu=3/2$ it gives $3\CC\cdot p^3j/3=\CC p^3j$, reproducing $\mathcal G_0$. For the auxiliary field, $2\Delta_\Phi-d=2$, and
\begin{equation}
 \mathcal G_\Phi(p)=\mathcal N_\Phi L^2\,p^2\ln p+a\,p^2=\frac{\mathcal N_\Phi L^2}{2}\,p^2\ln\frac{p^2}{\mu^2},
\end{equation}
where the arbitrary contact coefficient $a$ has been traded for a scheme scale $\mu$.

\emph{Step 3: position space.} Equation~\eqref{eq:FTpower} gives $c(1)=0$, because $1/\Gamma(-\nu)$ vanishes at $\nu=1$; this is the statement that $p^2$ alone is a contact term. Near $\nu=1+\varepsilon$, $1/\Gamma(-1-\varepsilon)\simeq\varepsilon$, so $c(1+\varepsilon)\simeq4\,\Gamma(\tfrac52)\varepsilon/\pi^{3/2}=3\varepsilon/\pi$. Using $p^2\ln p=\tfrac12\partial_\nu p^{2\nu}$ at $\nu=1$, we find
\begin{equation}
 \int\frac{\dd^3p}{(2\pi)^3}e^{ip\cdot x}p^2\ln p=\frac12\,c'(1)\,r^{-5}=\frac{3}{2\pi r^5},
\end{equation}
while the $p^2\ln\mu^2$ term contributes only at $x=0$. The auxiliary correlator at separated points is therefore
\begin{equation}
 \mathcal G_\Phi(x)=\frac{3\,\mathcal N_\Phi L^2}{2\pi|x|^5}\qquad(x\neq0).
\end{equation}
The result is positive, as it must be for a unitary operator, and has the power $|x|^{-2\Delta_\Phi}$ required by conformal invariance. Its coefficient agrees with the general normalization
\begin{equation}
 \langle\OO(x)\OO(0)\rangle=\mathcal NL^2\,\frac{2\nu\,\Gamma(\Delta)}{\pi^{d/2}\Gamma(\nu)}\,\frac{1}{|x|^{2\Delta}},
\end{equation}
which also gives $12/\pi^2$ for $\nu=3/2$, in agreement with Eq.~\eqref{eq:ordinaryG}.

This construction has three limitations. First, the full functional \eqref{eq:Wsplit} includes the unsuppressed auxiliary response $\mathcal G_\Phi\sim\mathcal N_\Phi L^2p^2\ln p$, whereas the filtered probe response decays exponentially. Thus taking $\mathcal N_\Phi\ll\CC$ cannot suppress the auxiliary contribution uniformly in momentum. The desired functional $W_0[\PP_{\lz}J]$ is obtained by retaining only the probe response and treating the auxiliary field as prescribed, non-fluctuating transfer data. Sending its action normalization to zero is not by itself a controlled semiclassical limit of a dynamical auxiliary field.

Second, a simultaneous variation of both fields subject to the identification changes the auxiliary problem. In particular,
\begin{equation}
 \delta I_{\phi,\mathrm{ren}}
 =\int\dd^3x\,\frac{\delta I_{\phi,\mathrm{ren}}}{\delta j(x)}
 \lz^{-1/2}\delta\Phi(\lz,x).
\end{equation}
This contributes a term supported at $z=\lz$ to the auxiliary variational equation. The homogeneous solution used above therefore need not solve the coupled stationary-action problem. A dynamical realization requires a specified surface coupling or constraint action and the corresponding matching conditions; none is supplied here. Equation~\eqref{eq:Wsplit} denotes the sequentially prescribed composition, not the on-shell functional of a demonstrated coupled model.

Third, $\OO_{\lz}$ is smeared with the reduced $\Delta=5/2$ kernel, rather than the $\Delta=3$ kernel of $\OO$. This construction is not the HKLL reconstruction \cite{HKLL} of the probe field. The exact identity \eqref{eq:chi1} motivates a common radial surface, but it does not dynamically select the entropy anchor.

\section{Zero-point length as a bulk proper length: finite one-loop tadpoles}
\label{sec:bulkloops}
In the previous sections $\lz$ was a coordinate length on the boundary. We now take the second reading described in the Introduction and treat $\lz$ as an AdS-invariant proper length in the bulk, entering the bulk propagator through the same proper-time weight as in flat space. We work in units $L=1$ unless $L$ is displayed.

Let $K(s;X,X')$ be the heat kernel of $-\Box+m^2$ on Euclidean $\mathrm{AdS}_4\cong H^4$, defined by $\partial_sK=-(-\Box+m^2)K$ with $K\to\delta(X,X')$ as $s\to0$. The ordinary propagator is $\int_0^\infty\dd s\,K$. Following Eq.~\eqref{eq:propagator}, we define the zero-point-length propagator by inserting the weight $e^{-\lz^2/4s}$:
\begin{equation}
 G_{\lz}(X,X')=\int_0^\infty\dd s\,K(s;X,X')\,e^{-\lz^2/4s}.
 \label{eq:Gl0cov}
\end{equation}
On $H^4$ the heat kernel depends on its two arguments only through the geodesic distance $\sigma$, and therefore so does $G_{\lz}$. The modified propagator is invariant under all AdS isometries.

\emph{Covariant source.} To see what equation $G_{\lz}$ satisfies, act with the wave operator. Using the heat equation under the integral and integrating by parts in $s$,
\begin{equation}
 \begin{split}
 (-\Box+m^2)G_{\lz}&=\int_0^\infty\dd s\,(-\partial_sK)\,e^{-\lz^2/4s}\\
 &=\frac{\lz^2}{4}\int_0^\infty\frac{\dd s}{s^2}K(s;X,X')\,e^{-\lz^2/4s}\\
 &\equiv\rho^{\rm AdS}_{\lz}(\sigma).
 \end{split}
 \label{eq:rhoAdS}
\end{equation}
The boundary terms vanish: at $s\to0$ the factor $e^{-\lz^2/4s}$ kills the delta function, and at $s\to\infty$ the heat kernel decays. So $G_{\lz}$ is the field produced by a smooth distributed source $\rho^{\rm AdS}_{\lz}$ instead of a point source. Its total strength follows from stochastic completeness of the heat kernel on $H^4$, $\int\dd V\,K=e^{-m^2s}$:
\begin{equation}
 \begin{split}
 \int\dd V\,\rho^{\rm AdS}_{\lz}&=\frac{\lz^2}{4}\int_0^\infty\frac{\dd s}{s^2}e^{-m^2s-\lz^2/4s}\\
 &=\lz m\,K_1(\lz m)=\PP_{\lz}(m),
 \end{split}
\end{equation}
which equals one for $m=0$. This volume-integrated normalization is valid for $m^2\geq0$, with the massless value understood as a limit. For BF-stable negative $m^2$, the integrated heat kernel grows as $e^{|m^2|s}$ and the displayed total-strength integral diverges at large proper time. Pointwise propagators and coincidence integrals may nevertheless remain finite because of the AdS spectral gap. We do not interpret the tachyonic case as a finite-total-strength source or define its normalization by analytic continuation of this volume integral. With the flat kernel $(4\pi s)^{-2}e^{-x^2/4s}$, Eq.~\eqref{eq:rhoAdS} reduces to the four-dimensional kernel $\rho^{(4)}_{\lz}$ of Eq.~\eqref{eq:rho}. In AdS, the heat kernel approaches the flat one at short distances and short proper times, with corrections controlled by the curvature \cite{Vassilevich,GrigoryanNoguchi}, so that $\rho^{\rm AdS}_{\lz}=\rho^{(4)}_{\lz}\left[1+O(\lz^2/L^2,\sigma^2/L^2)\right]$ for $\lz,\sigma\ll L$. Equation~\eqref{eq:rhoAdS} is therefore the AdS-covariant distributed source called for in Sec.~\ref{sec:extensions}.

In flat space, Eq.~\eqref{eq:Gl0cov} is equivalent to Padmanabhan's replacement $\sigma^2\to\sigma^2+\lz^2$ in the propagator. In AdS the heat kernel is not a pure Gaussian in $\sigma$, and the two definitions differ. We use Eq.~\eqref{eq:Gl0cov} because it is manifestly covariant and keeps the proper-time origin of the zero-point length.

\emph{Consequence for the boundary.} Dependence on $\sigma$ ensures AdS invariance of the prescribed bulk two-point kernel. If the action, interactions, boundary conditions and renormalization prescription also preserve these isometries, the extrapolated boundary two-point function retains the conformal form, with dimensions determined by its asymptotic behavior and quantum corrections. Invariance of a propagator alone does not establish conformal Ward identities for every interacting correlator. A fixed bulk proper length introduces a dimensionless ratio $\lz/L$, rather than a boundary coordinate cutoff, and does not by itself soften the boundary short-distance power law. This is why the boundary resolution of Secs.~\ref{sec:kernel}--\ref{sec:entropy} had to be a coordinate length that breaks dilatations (Sec.~\ref{sec:extensions}). What a bulk proper length can do is regulate bulk loops, and we now compute the simplest example.

We now turn to the coincidence limit. The one-loop tadpole requires the propagator at coincident points, $G_{\lz}(X,X)$. Without the zero-point length this quantity diverges as $1/\sigma^2$. To compute it we use the spectral decomposition of the Laplacian on $H^4$. The spectrum of $-\Box$ is continuous, $\lambda^2+\tfrac94$ with $\lambda\geq0$, and the Plancherel density is proportional to $\lambda(\lambda^2+\tfrac14)\tanh\pi\lambda$ \cite{Camporesi}. At large $\lambda$ this density grows as $\lambda^3$, as in flat four-dimensional space, and we fix its normalization by matching to flat space, $\frac{1}{8\pi^2}\int_0^\infty\dd\lambda\,\lambda^3e^{-s\lambda^2}=(4\pi s)^{-2}$. Writing $\nu^2=m^2+\tfrac94$, the heat kernel at coincident points is
\begin{equation}
 \begin{split}
 K(s;X,X)={}&\frac{1}{8\pi^2}\int_0^\infty\dd\lambda\,\lambda\left(\lambda^2+\tfrac14\right)\\
 &\times\tanh(\pi\lambda)\,e^{-s(\lambda^2+\nu^2)} .
 \end{split}
\end{equation}
Inserting this into Eq.~\eqref{eq:Gl0cov}, the proper-time integral for each mode is the flat-space integral \eqref{eq:propagator} with $p\to\omega$, $\int_0^\infty\dd s\,e^{-s\omega^2-\lz^2/4s}=(\lz/\omega)K_1(\lz\omega)$, where $\omega=\sqrt{\lambda^2+\nu^2}$. The exact coincidence value is therefore
\begin{equation}
 \begin{split}
 G_{\lz}(X,X)={}&\frac{1}{8\pi^2}\int_0^\infty\dd\lambda\,\lambda\left(\lambda^2+\tfrac14\right)\\
 &\times\tanh(\pi\lambda)\,\frac{\lz}{\omega}K_1(\lz\omega).
 \end{split}
 \label{eq:Gspectral}
\end{equation}
At large $\lambda$ the factor $K_1(\lz\omega)$ decays exponentially, so the integral converges for any $\lz>0$.

To evaluate Eq.~\eqref{eq:Gspectral} for small $\lz$ we split $\tanh\pi\lambda=1-2/(e^{2\pi\lambda}+1)$ and write $G_{\lz}(X,X)=A+B$. The first part contains the ultraviolet behavior and can be done exactly; the second is ultraviolet finite and encodes the curvature of AdS.

\emph{The $A$ part.} Changing variable to $u=\omega$, with $\lambda\,\dd\lambda=u\,\dd u$ and $\lambda^2+\tfrac14=u^2-(\nu^2-\tfrac14)$, gives
\begin{equation}
 A=\frac{\lz}{8\pi^2}\int_\nu^\infty\dd u\left[u^2-\left(\nu^2-\tfrac14\right)\right]K_1(\lz u).
\end{equation}
Both terms are elementary Bessel integrals. Using $\frac{\dd}{\dd x}[x^2K_2(x)]=-x^2K_1(x)$ and $\frac{\dd}{\dd x}K_0(x)=-K_1(x)$,
\begin{equation}
 A=\frac{1}{8\pi^2}\left[\nu^2K_2(\nu\lz)-\left(\nu^2-\tfrac14\right)K_0(\nu\lz)\right],
\end{equation}
which is exact. For small $\lz$ we use $K_2(x)=\frac{2}{x^2}-\frac12+O(x^2\ln x)$ and $K_0(x)=-\ln\frac x2-\gamma_E+O(x^2\ln x)$:
\begin{equation}
 A=\frac{1}{4\pi^2\lz^2}+\frac{\nu^2-\frac14}{8\pi^2}\left[\ln\frac{\nu\lz}{2}+\gamma_E\right]-\frac{\nu^2}{16\pi^2}+O(\lz^2\ln\lz).
\end{equation}
The leading term $1/(4\pi^2\lz^2)$ is the flat-space value. It is the finite remnant of the $1/\sigma^2$ divergence, cut off at $\sigma\sim\lz$.

\emph{The $B$ part.} The factor $1/(e^{2\pi\lambda}+1)$ restricts this part to $\lambda\lesssim1$, that is, to modes on the curvature scale. Its limit $\lz\to0$ is therefore finite and is obtained by replacing $(\lz/\omega)K_1(\lz\omega)$ with $1/\omega^2$:
\begin{equation}
 B=-\frac{1}{4\pi^2}\int_0^\infty\dd\lambda\,\frac{\lambda(\lambda^2+\frac14)}{(\lambda^2+\nu^2)(e^{2\pi\lambda}+1)}+O(\lz^2\ln\lz).
\end{equation}
Partial fractions give $\frac{\lambda(\lambda^2+1/4)}{\lambda^2+\nu^2}=\lambda-\frac{(\nu^2-\frac14)\lambda}{\lambda^2+\nu^2}$, which leaves two integrals. The first is a standard Fermi-type integral:
\begin{equation}
 \int_0^\infty\frac{\lambda\,\dd\lambda}{e^{2\pi\lambda}+1}=\frac{1}{4\pi^2}\int_0^\infty\frac{x\,\dd x}{e^x+1}=\frac{1}{4\pi^2}\cdot\frac{\pi^2}{12}=\frac{1}{48}.
\end{equation}
For the second we convert the Fermi factor into Bose factors, $\frac{1}{e^{x}+1}=\frac{1}{e^x-1}-\frac{2}{e^{2x}-1}$, and use Binet's second formula,
\begin{equation}
 \mathcal J(z)\equiv\int_0^\infty\frac{t\,\dd t}{(t^2+z^2)(e^{2\pi t}-1)}=\frac12\left[\ln z-\frac1{2z}-\psi(z)\right].
\end{equation}
After the rescaling $t\to t/2$, the second Bose term becomes $\mathcal J(2\nu)$. The duplication formula $\psi(2\nu)=\frac12\psi(\nu)+\frac12\psi(\nu+\frac12)+\ln2$ then gives
\begin{equation}
 \begin{split}
 \int_0^\infty\frac{\lambda\,\dd\lambda}{(\lambda^2+\nu^2)(e^{2\pi\lambda}+1)}&=\mathcal J(\nu)-2\mathcal J(2\nu)\\
 &=\frac12\left[\psi\!\left(\nu+\tfrac12\right)-\ln\nu\right],
 \end{split}
\end{equation}
in which the $1/(2z)$ terms of Binet's formula cancel. Collecting the pieces,
\begin{equation}
 B=-\frac{1}{4\pi^2}\left[\frac{1}{48}-\frac{\nu^2-\frac14}{2}\left(\psi\!\left(\nu+\tfrac12\right)-\ln\nu\right)\right].
\end{equation}

\emph{Result.} In $A+B$ the terms in $\ln\nu$ cancel, leaving $\ln\lz$ as the only logarithm. We express the result in terms of $\Delta=\tfrac32+\nu$, using $\nu^2-\tfrac14=(\Delta-1)(\Delta-2)$ and $\nu+\tfrac12=\Delta-1$, and restore $L$:
\begin{multline}
 G_{\lz}(X,X)=\frac{1}{4\pi^2\lz^2}+\frac{(\Delta-1)(\Delta-2)}{8\pi^2L^2}\ln\frac{\lz}{L}+\frac{c_\Delta}{L^2}\\
 +O\!\left(\frac{\lz^2}{L^4}\ln\frac{\lz}{L}\right),
 \label{eq:Gcoinc}
\end{multline}
with $\nu=\Delta-\tfrac32$ and the finite constant
\begin{multline}
 c_\Delta=\frac{1}{8\pi^2}\Big[(\Delta-1)(\Delta-2)\big(\psi(\Delta-1)+\gamma_E-\ln2\big)\\
 -\frac{\nu^2}{2}-\frac{1}{24}\Big].
 \label{eq:cDelta}
\end{multline}
The three terms in Eq.~\eqref{eq:Gcoinc} have distinct origins. The first is the flat-space short-distance behavior, the logarithmic coefficient combines the mass and curvature through $m^2+2/L^2$, and the constant $c_\Delta$ depends on both the curvature and the specific form of the zero-point-length regulator.

The result \eqref{eq:Gcoinc} passes three independent analytic checks.
\begin{enumerate}[leftmargin=*,itemsep=2pt]
 \item \emph{Heat-kernel coefficient.} At short proper time, $K(s;X,X)=(4\pi s)^{-2}e^{-m^2s}\left[1+a_1s+\dots\right]$. The $s^{-1}$ term of the expansion produces the logarithm in $G_{\lz}(X,X)$, with coefficient $(m^2-a_1)/(8\pi^2)$. For a minimally coupled scalar $a_1=R/6$, and on $H^4$ with $R=-12/L^2$ this is $a_1=-2/L^2$ \cite{Vassilevich}. Since $m^2L^2+2=(\Delta-1)(\Delta-2)$, the coefficient agrees with Eq.~\eqref{eq:Gcoinc}.
 \item \emph{Flat-space limit.} For $\nu\to\infty$ with $m\simeq\nu/L$ fixed, the curvature becomes negligible on the scale of the Compton wavelength. Using $\psi(\nu+\frac12)\simeq\ln\nu$, Eq.~\eqref{eq:Gcoinc} becomes
 \begin{equation*}
 \frac{1}{4\pi^2\lz^2}+\frac{m^2}{8\pi^2}\left[\ln\frac{m\lz}{2}+\gamma_E-\frac12\right],
 \end{equation*}
 which is the small-$\lz$ expansion of the flat-space result $mK_1(m\lz)/(4\pi^2\lz)$.
 \item \emph{Conformal coupling, $\Delta=2$.} Here the logarithm vanishes, and Eq.~\eqref{eq:cDelta} gives $c_2=-1/(48\pi^2)$. The exact propagator is known in closed form, $G_{\Delta=2}=\frac{1}{8\pi^2}\left(\frac1u-\frac1{u+2}\right)$ with $u=\cosh\sigma-1$. Splitting points gives $\frac{1}{4\pi^2\sigma^2}-\frac{1}{12\pi^2}$. The two regularizations differ by the Van Vleck factor $\Delta_{\rm VV}^{1/2}=(\sigma/\sinh\sigma)^{3/2}=1-\sigma^2/4+\dots$, which multiplies the leading $1/(4\pi^2\sigma^2)$ term in point splitting and shifts the constant by $-1/(16\pi^2)$. Indeed $-\frac{1}{48\pi^2}-\frac{1}{16\pi^2}=-\frac{1}{12\pi^2}$.
\end{enumerate}

The coincidence value enters physical quantities through loops. Add a quartic interaction $V=\frac{\lambda}{4!}\varphi^4$ for the canonically normalized field $\varphi=\sqrt{\mathcal N}\phi$. At one loop, the tadpole diagram shifts the mass by
\begin{equation}
 \delta m^2=\frac{\lambda}{2}\,G_{\lz}(X,X).
\end{equation}
The factor $\frac12$ is the symmetry factor of the tadpole: $\frac{\lambda}{4!}$ times the $\binom42=6$ ways of choosing the contracted pair, times $2$ for the assignment of the external legs. Since AdS is homogeneous, $G_{\lz}(X,X)$ is the same at every point and the shift is a constant. In holography a mass shift is a shift of the scaling dimension. Varying $m^2L^2=\Delta(\Delta-3)$ gives $\delta(m^2L^2)=(2\Delta-3)\delta\Delta=2\nu\,\delta\Delta$, so the anomalous dimension is
\begin{equation}
 \begin{split}
 \gamma\equiv\delta\Delta={}&\frac{\lambda L^2}{4\nu}G_{\lz}(X,X)\\
 ={}&\frac{\lambda}{16\pi^2\nu}\frac{L^2}{\lz^2}+\frac{\lambda(\Delta-1)(\Delta-2)}{32\pi^2\nu}\ln\frac{\lz}{L}\\
 &+\frac{\lambda c_\Delta}{4\nu}.
 \end{split}
\end{equation}
For the massless probe, $\Delta=3$ and $\nu=3/2$. With $\psi(2)=1-\gamma_E$, Eq.~\eqref{eq:cDelta} gives $c_3=\frac{1}{8\pi^2}\left(\frac56-2\ln2\right)\approx-7.003\times10^{-3}$, and
\begin{equation}
 \begin{split}
 \gamma_{\Delta=3}={}&\frac{\lambda}{24\pi^2}\left[\frac{L^2}{\lz^2}+\ln\frac{\lz}{L}+\frac{5}{12}-\ln2\right]\\
 &+O\!\left(\lambda\frac{\lz^2}{L^2}\ln\frac{\lz}{L}\right).
 \end{split}
 \label{eq:gamma3}
\end{equation}
This regulated one-loop mass correction is finite. Its interpretation as an anomalous dimension uses the pole connected to the original scalar and perturbative control of the mass shift. A physical prediction further requires a specified renormalized mass condition and microscopic matching; the finite regulator alone does not fix these inputs.

Three features of this result deserve comment.
\begin{enumerate}[leftmargin=*,itemsep=2pt]
 \item \emph{Universal versus regulator-specific content.} The coefficient of $\ln(\lz/L)$ is fixed by $a_1$ and would be the same for any regulator. The zero-point length determines the power-law term and the constant $c_\Delta$, and the latter differs from the point-splitting constant, as check 3 shows. In standard holography the $L^2/\lz^2$ term is absorbed into the bare mass so as to hold $\Delta$ fixed. It becomes a prediction only if a microscopic theory fixes both $\lz$ and the bare mass.
 \item \emph{Control of the $1/N$ expansion.} If the coupling scales as $\lambda=O(G_N/L^2)$, the leading term is $\gamma\sim G_N/\lz^2\sim\ell_P^2/\lz^2$. Perturbation theory therefore requires $\lz\gg\ell_P$, as would hold for $\lz\sim\sqrt{\alpha'}$ at weak string coupling. A zero-point length at the Planck scale would make the one-loop shift of order one.
 \item \emph{No boundary regulation.} At this order the boundary two-point function is $|x|^{-2(\Delta+\gamma)}$. The bulk zero-point length shifts $\Delta$ but does not regulate boundary short-distance behavior. The boundary reading of Secs.~\ref{sec:kernel}--\ref{sec:entropy} and the bulk reading of this section do different jobs and complement each other.
\end{enumerate}

\section{Spectral test of holographic consistency}
\label{sec:spectral}
No single calculation can show that a bulk construction has a dual. What can be tested is \emph{holographic consistency}: whether the boundary data the construction produces satisfy the requirements of a conformal field theory, namely conformal covariance, a sensible operator spectrum and unitarity. In this section we carry out the test for the AdS-covariant zero-point-length propagator \eqref{eq:Gl0cov}. The tool is the K\"all\'en--Lehmann decomposition, which writes a bulk two-point function as a sum of ordinary AdS propagators and reads off the boundary spectrum and its weights. We work in $\mathrm{AdS}_{d+1}$ with $L=1$, restoring $L$ through $\lz\to\lz/L$ and $M\to ML$ at the end.

The natural basis for functions on AdS that depend only on geodesic distance is the set of harmonic functions $\Omega_\lambda(X,Y)$, $\lambda\in\mathbb R$, which play the role of plane waves \cite{Penedones}. They are eigenfunctions of the Laplacian,
\begin{equation}
 \left(\Box+\tfrac{d^2}{4}+\lambda^2\right)\Omega_\lambda=0,
\end{equation}
they can be written as a difference of propagators with complex dimension,
\begin{equation}
 \Omega_\lambda=\frac{i\lambda}{2\pi}\left(G_{\frac d2+i\lambda}-G_{\frac d2-i\lambda}\right),
\end{equation}
and they form a complete set,
\begin{equation}
 \int_{-\infty}^{\infty}\dd\lambda\,\Omega_\lambda(X,Y)=\frac{\delta(X,Y)}{\sqrt g}.
\end{equation}
Here $G_\Delta$ is the standard bulk-to-bulk propagator,
\begin{equation}
 \begin{split}
 G_\Delta(\xi)={}&\frac{\Gamma(\Delta)\,\xi^\Delta}{2^{\Delta+1}\pi^{d/2}\Gamma(\Delta-\frac d2+1)}\\
 &\times{}_2F_1\!\left(\tfrac\Delta2,\tfrac{\Delta+1}2;\Delta-\tfrac d2+1;\xi^2\right),
 \end{split}
 \label{eq:GDelta}
\end{equation}
with $\xi=1/\cosh\sigma$.
On $\Omega_\lambda$ the operator $-\Box+m^2$ acts as multiplication by $\lambda^2+\frac{d^2}{4}+m^2=\lambda^2+\nu^2$. Any function of $-\Box+m^2$ therefore becomes the same function of $\lambda^2+\nu^2$ in this basis. In particular, the ordinary propagator, which inverts $-\Box+m^2$, is
\begin{equation}
 G_\Delta=\int_{-\infty}^{\infty}\dd\lambda\,\frac{\Omega_\lambda}{\lambda^2+\nu^2},
\end{equation}
and the heat kernel, which exponentiates it, is
\begin{equation}
 K(s;X,Y)=\int_{-\infty}^{\infty}\dd\lambda\,\Omega_\lambda\,e^{-s(\lambda^2+\nu^2)} .
\end{equation}
Inserting the heat-kernel representation into Eq.~\eqref{eq:Gl0cov} and doing the $s$-integral mode by mode gives
\begin{equation}
 G_{\lz}(X,Y)=\int_{-\infty}^{\infty}\dd\lambda\;\Omega_\lambda(X,Y)\;\frac{\lz}{\omega}K_1(\lz\omega),
 \label{eq:Gharmonic}
\end{equation}
with $\omega^2=\lambda^2+\nu^2$. In the AdS spectral variable, the zero-point-length propagator is exactly Padmanabhan's flat-space form factor \eqref{eq:propagator}, with $p$ replaced by $\omega$. The ordinary propagator corresponds to $1/\omega^2$.

Equation~\eqref{eq:Gharmonic} is not yet a sum of propagators, because $(\lz/\omega)K_1(\lz\omega)$ is not of the form $1/(\omega^2+M^2)$. We now write it as a superposition of such terms. Two Hankel-type integrals are needed. The first is elementary,
\begin{equation}
 \int_0^\infty J_1(bt)\,\dd t=\frac1b,
\end{equation}
and the second is a standard Hankel transform,
\begin{equation}
 \int_0^\infty J_1(bt)\,\frac{t^2}{t^2+a^2}\,\dd t=a\,K_1(ab),
 \qquad a,b>0 .
\end{equation}

\emph{Step 1.} Decompose $\frac{1}{t^2+a^2}=\frac{1}{a^2}\left(1-\frac{t^2}{t^2+a^2}\right)$ and integrate against $J_1(bt)$:
\begin{equation}
 \int_0^\infty\frac{J_1(bt)}{t^2+a^2}\,\dd t=\frac{1}{a^2}\left[\frac1b-a\,K_1(ab)\right].
\end{equation}

\emph{Step 2.} Set $b=\lz$, $a=\omega$ and $t=M$, multiply by $\lz$ and rearrange:
\begin{equation}
 \frac{\lz}{\omega}K_1(\lz\omega)=\frac{1}{\omega^2}-\lz\int_0^\infty\dd M\,\frac{J_1(\lz M)}{\omega^2+M^2}.
 \label{eq:dispersion}
\end{equation}
The zero-point-length form factor is the massless propagator $1/\omega^2$ minus a superposition of massive ones, $1/(\omega^2+M^2)$, weighted by $\lz J_1(\lz M)$.

\emph{Step 3.} With the measure $\dd M^2=2M\,\dd M$, Eq.~\eqref{eq:dispersion} takes the form of a K\"all\'en--Lehmann representation in $\omega^2$,
\begin{equation}
 \frac{\lz}{\omega}K_1(\lz\omega)=\int_0^\infty\dd M^2\,\frac{\rho(M^2)}{\omega^2+M^2},
 \label{eq:KLflat}
\end{equation}
with the spectral density
\begin{equation}
 \rho(M^2)=\delta(M^2)+\rho_c(M^2),
 \qquad
 \rho_c(M^2)=-\frac{\lz\,J_1(\lz M)}{2M}.
 \label{eq:rhoc}
\end{equation}
The spectral density has a pole of unit weight at $M^2=0$ and a continuum $\rho_c$. The same density follows from the discontinuity across the branch cut of $K_1$. Using $K_1(ix)=-\frac\pi2\left[J_1(x)-iY_1(x)\right]$ at $\omega^2=-M^2+i0$, one finds $\operatorname{Im}\left[\frac{\lz}{\omega}K_1(\lz\omega)\right]=\frac{\pi\lz}{2M}J_1(\lz M)=-\pi\rho_c(M^2)$. No subtraction is needed, because $\frac{\lz}{\omega}K_1(\lz\omega)$ decays as $|\omega|^{-3/2}$ throughout the cut $\omega^2$-plane.

\emph{Step 4: sum rule.} Since $\int_0^\infty J_1(x)\,\dd x=1$, the total continuum weight is
\begin{equation}
 \int_0^\infty\dd M^2\,\rho_c(M^2)=-\lz\int_0^\infty\dd M\,J_1(\lz M)=-1 .
 \label{eq:sumrule}
\end{equation}
The total spectral weight, pole plus continuum, is therefore $1-1=0$. This value is forced. At large $\omega$ the representation \eqref{eq:KLflat} behaves as $\omega^{-2}\int\rho$, while the left-hand side decays faster than $\omega^{-2}$, so $\int\rho$ must vanish. In other words, the softness of the zero-point-length propagator in the ultraviolet requires the continuum to cancel the pole exactly, and no positive density can do that.

\emph{Step 5: cumulative weight.} Using $\frac{\dd}{\dd x}J_0(x)=-J_1(x)$, the continuum weight accumulated up to a mass $M$ is
\begin{equation}
 \begin{split}
 W(M)&\equiv\int_0^{M^2}\dd M'^2\,\rho_c(M'^2)=-\left[1-J_0(\lz M)\right]\\
 &\simeq-\frac{(\lz M)^2}{4}\qquad(\lz M\ll1).
 \end{split}
 \label{eq:cumulative}
\end{equation}
For $\lz M\ll1$ the negative weight is small, of order $(\lz M)^2$. It approaches the full value $-1$, with oscillations, only for $\lz M\gtrsim1$.

With this dispersion relation in hand, we return to AdS. Inserting Eq.~\eqref{eq:dispersion} into Eq.~\eqref{eq:Gharmonic}, each term $1/(\lambda^2+\nu^2+M^2)$ is the spectral function of an ordinary propagator with $\nu_M^2=\nu^2+M^2$. The zero-point-length propagator is therefore a superposition of ordinary AdS propagators:
\begin{equation}
 \begin{split}
 G_{\lz}(X,Y)={}&G_\Delta(X,Y)\\
 &-\lz\int_0^\infty\dd M\,J_1(\lz M)\,G_{\Delta(M)}(X,Y).
 \end{split}
 \label{eq:KLAdS}
\end{equation}
Each $G_{\Delta(M)}$ is the propagator of a field dual to a boundary operator of dimension
\begin{equation}
 \Delta(M)=\frac d2+\sqrt{\nu^2+M^2}.
 \label{eq:DeltaM}
\end{equation}
Changing variables from $M$ to this dimension, $\Delta'=\Delta(M)$, with $M\,\dd M=(\Delta'-\frac d2)\,\dd\Delta'$, gives
\begin{equation}
 G_{\lz}=G_\Delta+\int_\Delta^\infty\dd\Delta'\,\varrho(\Delta')\,G_{\Delta'},
 \label{eq:KLDelta}
\end{equation}
with the spectral density
\begin{equation}
 \varrho(\Delta')=-\lz\,\frac{J_1(\lz M)}{M}\left(\Delta'-\frac d2\right),
 \label{eq:varrho}
\end{equation}
where $M=\sqrt{(\Delta'-\frac d2)^2-\nu^2}$ inverts Eq.~\eqref{eq:DeltaM}. The weight $\varrho(\Delta')$ is the spectral density in terms of boundary dimensions. Near threshold $J_1(\lz M)/M\to\lz/2$, so the density approaches a constant,
\begin{equation}
 \varrho(\Delta')\to-\frac{\lz^2}{2}\,\nu\qquad(\Delta'\to\Delta^+),
\end{equation}
which is negative. The weight changes sign at the zeros $j_{1,k}$ of $J_1$, that is, at the dimensions
\begin{equation}
 \Delta'_k=\frac d2+\sqrt{\nu^2+j_{1,k}^2\,\frac{L^2}{\lz^2}}\;\simeq\;\frac d2+j_{1,k}\,\frac{L}{\lz},
 \label{eq:signchanges}
\end{equation}
with $j_{1,k}=3.8317,\ 7.0156,\ 10.1735,\ \dots$. For $d=3$ and $\nu=3/2$, for example, $\lz/L=0.1$ gives $\Delta'_k\simeq39.9,\ 71.7,\ 103.3$, and $\lz/L=0.01$ gives $\Delta'_k\simeq384.7,\ 703.1,\ 1018.9$.

To interpret Eq.~\eqref{eq:KLAdS} we compare it with what a unitary holographic theory requires. Suppose $\phi$ is a bulk field in a theory with a unitary boundary CFT. Energy eigenstates in global AdS are CFT states, organized into conformal families labelled by their dimensions $\Delta'$. Inserting a complete set of such states into the bulk two-point function, and using AdS covariance to resum each family, gives \cite{Penedones}
\begin{equation}
 \langle\phi(X)\phi(Y)\rangle=\sum_{\Delta'}\big|\langle0|\phi(0)|\Delta'\rangle\big|^2\,G_{\Delta'}(X,Y).
 \label{eq:KLcriterion}
\end{equation}
The weights $|\langle0|\phi(0)|\Delta'\rangle|^2$ are squared norms of matrix elements, so unitarity requires them to be non-negative. Comparing Eq.~\eqref{eq:KLcriterion} with Eqs.~\eqref{eq:KLDelta} and \eqref{eq:varrho} leads to three conclusions.
\begin{enumerate}[leftmargin=*,itemsep=3pt]
 \item \emph{Conformal symmetry is preserved.} Every term in Eq.~\eqref{eq:KLAdS} is an AdS-covariant propagator $G_{\Delta'}$, so the kernel itself is AdS invariant. Conformal covariance of interacting boundary correlators additionally requires invariant interactions, boundary conditions and renormalization. For the free kernel, the original pole controls the leading extrapolate boundary limit. The continuum starts at the same threshold and has no finite gap above $\Delta$, so its integrated suppression need not be a fixed extra power of the radial coordinate; for $\nu>0$ its finite threshold density gives logarithmically suppressed corrections after the leading boundary power is stripped off. The leading extrapolated two-point function retains the original dimension and pole weight. This free-kernel statement does not establish a full unitary single-trace spectrum.
 \item \emph{The signed decomposition contains a continuum.} Besides the original pole, the mathematical decomposition contains propagators labelled by $\Delta'\in(\Delta,\infty)$. Because their weights are not nonnegative, this cannot be read as an established spectrum of physical operators in a unitary boundary CFT.
 \item \emph{Unitarity is violated.} The continuum weight \eqref{eq:varrho} is negative at threshold and oscillates in sign, with total weight $-1$ by Eq.~\eqref{eq:sumrule}. Hence $G_{\lz}$ cannot be the two-point function of a bulk field in a unitary AdS/CFT. This is the bulk counterpart of the failure of reflection positivity, Eq.~\eqref{eq:RPfail}, for the boundary filter.
\end{enumerate}

The size of a low-spectral-scale modification can be quantified without interpreting it as restoration of unitarity. Restoring dimensions and writing $\epsilon=\lz/L$ and $\widehat M=LM$, the integrated signed continuum weight below $\widehat M_{\max}$ is
\begin{equation}
 W(\widehat M_{\max})=J_0(\epsilon\widehat M_{\max})-1
 =-\frac{\epsilon^2\widehat M_{\max}^2}{4}
 +O(\epsilon^4\widehat M_{\max}^4).
\end{equation}
For a boundary dimension cutoff $\Delta_{\max}$,
\begin{equation}
 \widehat M_{\max}^2=
 (\Delta_{\max}-d/2)^2-\nu^2.
\end{equation}
At fixed $\Delta_{\max}$ this signed weight is $O(\epsilon^2)$, but the estimate is not uniform as $\Delta_{\max}$ increases toward $\epsilon^{-1}$. Near that scale the modification becomes order one. Beyond the first zero of $J_1$, $W$ is a signed cumulative weight, not the total magnitude of negative spectral weight.

The continuum remains negative arbitrarily close to threshold for every $\lz>0$. Positivity is an exact necessary condition, so a parametrically small violation does not establish a unitary low-energy theory. Nor does a bound on this integrated weight alone bound errors in every observable. The bulk construction is therefore a covariant Euclidean regulator with a demonstrable positivity obstruction when its exact kernel is interpreted as a physical propagator. A unitary Wilsonian interpretation would require separate matching and positivity arguments, which are not provided here.

The sum rule \eqref{eq:sumrule} makes the obstruction especially transparent: a positive unit pole and a nonnegative continuum cannot have zero total spectral weight. The oscillatory continuum integral is conditionally convergent; its value follows from the stated Bessel integral, rather than an unrestricted exchange of a large-$\omega$ expansion and the spectral integral. A high-mass positive tower added without changing the low-mass density cannot remove the already negative discontinuity near threshold. Any unitary completion must alter the spectral content of the prescribed propagator, or change its interpretation as the correlator of a physical bulk field.

The scale $\Delta'\sim L/\lz$ and the sign-change dimensions in Eq.~\eqref{eq:signchanges} characterize this particular form factor. They do not prove that a dual CFT exists or predict a mandatory tower of positive-norm operators. If $\lz\sim\sqrt{\alpha'}$, the parametric scale agrees with the familiar massive-string scale in AdS \cite{GKP}; this comparison is suggestive but supplies neither a string derivation nor positivity. Our spectral result is an obstruction and a constraint on possible completions, not a demonstration of holographic consistency.

\section{Conclusions}
\label{sec:conclusions} We have studied the zero-point length
$\lz$ of T-duality-inspired effective field theory in AdS/CFT, in
two readings. In the first, $\lz$ is a resolution scale for
boundary measurements. Imposing the Bessel filter on the source of
a massless scalar in $\mathrm{AdS}_4$ gives the exact bulk
solution \eqref{eq:mainbulk} and the filtered two-point function
\eqref{eq:mainG}. Regularity fixes the bulk solution, a local
counterterm removes the radial divergence, and neither the bulk
mass nor the dimension of the underlying operator changes. No
independent momentum cutoff is needed: all momentum moments of the
response converge, and the coincidence value
\eqref{eq:coincidence} is finite. For the $\Delta=3$ probe this
value is fixed within the scale-covariant parent renormalization
prescription. At large separations the filtered correlator
approaches the conformal one with the logarithmic correction
\eqref{eq:largeR}, which reflects the power-law tails of the
zero-point-length kernel. Full Euclidean smearing fails reflection
positivity as a time-local covariance, while spatial smearing
preserves the parent spectral measure but leaves high frequencies
unregulated. A finite Euclidean response should therefore not be
mistaken for a unitary ultraviolet completion.

The filter has a direct holographic meaning. It is exactly the
bulk-to-boundary propagator of a $\nu=1$ field evaluated at depth
$z=\lz$, Eq.~\eqref{eq:chi1}, and an auxiliary two-field transfer
prescription represents the source map, without providing a
coupled dynamical realization. Using the same surface $z=\lz$ to
anchor Ryu--Takayanagi surfaces gives the exact disk entropy
\eqref{eq:diskentropy}. It is finite, reproduces the perimeter law
and the universal constant $-F$ at large radius, and vanishes
smoothly for subresolution regions. The $\mathrm{AdS}_3$ interval
entropy \eqref{eq:intervalentropy} follows in the same way. The
same prescription also governs correlations between two intervals.
For arbitrary interval lengths it yields the exact critical
separation \eqref{eq:MIcriticalunequalExact}, which reduces to
Eq.~\eqref{eq:MIcritical} for equal intervals. In the
equal-interval case, finite resolution moves the connectivity
transition to smaller separations. For arbitrary intervals the
transition exists and is unique, and the concavity of the
single-interval entropy implies that equal intervals maximize both
the mutual information and the connectivity window at fixed total
length, Eqs.~\eqref{eq:MIinfobound} and
\eqref{eq:MIasymmetrybound}. For subresolution intervals the
connected window shrinks as $h_c\simeq D_1D_2(D_1+D_2)/(16\lz^2)$,
Eq.~\eqref{eq:MIcriticalunequalUV}; asymmetry reduces it by the
factor $1-\delta^2$, and the rescaled mutual information collapses
onto the universal piecewise-linear profile \eqref{eq:MIscaling}.
These are statements about the classical geometric prescription
and not about entropies of the filtered observables. The entropy
prescription remains an assumption in addition to the source
filter, although both now refer to the same radial surface. In the
second reading, $\lz$ is an AdS-invariant bulk proper length.
Under an invariant bulk setup it preserves the conformal form of
boundary two-point functions while regulating the one-loop
tadpole. Equation~\eqref{eq:Gspectral} is the exact coincidence
integral; Eqs.~\eqref{eq:Gcoinc}--\eqref{eq:cDelta} give its
small-$\lz/L$ expansion, and Eq.~\eqref{eq:gamma3} gives the
associated perturbative pole shift subject to a
mass-renormalization prescription. The exact signed
K\"all\'en--Lehmann decomposition retains the original pole but
has negative continuum weight immediately above threshold. It
therefore fails the positivity condition required of a physical
bulk two-point function in a unitary holographic theory. Low-scale
signed corrections can be small, but this does not establish a
unitary effective theory. The scale $L/\lz$ is a crossover scale
of the form factor; a unitary completion and its operator spectrum
remain undetermined.

Several questions remain open. A dynamical origin for the surface
$z=\lz$, such as a defect or brane with its own action, would
replace the two prescriptions of the boundary reading with a
single bulk object. A boundary replica construction could decide
whether the source filter and the entropy anchoring descend from
one microscopic mechanism, and whether the geometric connectivity
threshold has a counterpart in the correlations of the smeared
observables. Finally, a string embedding that fixes $\lz$ in terms
of $\sqrt{\alpha'}$ and a compactification scale would turn the
present effective prescription into a derived one, and would test
whether a modified, positive spectral representation can arise
from the massive string spectrum while matching the desired
low-energy observables.

\end{document}